\documentclass[review,preprint,12pt]{elsarticle}

\usepackage{amsmath,amssymb,bm,graphicx,subfigure,enumerate,latexsym,comment,dcolumn}

\journal{}

\begin{document}

\begin{frontmatter}



	\title{Monte Carlo simulation for kinetic chemotaxis model: an application to the traveling population wave}


	\author{Shugo YASUDA\corref{cor1}}
	\address{Graduate School of Simulation Studies, University of Hyogo}
	\address{WPI Immunology Frontier Research Center, Osaka University}
	\cortext[cor1]{Email: yasuda@sim.u-hyogo.ac.jp, Tel/Fax +81(0)783031990}

	\begin{abstract}
		A Monte Carlo simulation of chemotactic bacteria is developed on the basis of the kinetic model and is applied to a one-dimensional traveling population wave in a microchannel. In this simulation, the Monte Carlo method, which calculates the run-and-tumble motions of bacteria, is coupled with a finite volume method to calculate the macroscopic transport of the chemical cues in the environment.
		The simulation method can successfully reproduce the traveling population wave of bacteria that was observed experimentally and reveal the microscopic dynamics of bacterium coupled with the macroscopic transports of the chemical cues and bacteria population density.
		The results obtained by the Monte Carlo method are also compared with the asymptotic solution derived from the kinetic chemotaxis equation in the continuum limit, where the Knudsen number, which is defined by the ratio of the mean free path of bacterium to the characteristic length of the system, vanishes.
		The validity of the Monte Carlo method in the asymptotic behaviors for small Knudsen numbers is numerically verified. 
	\end{abstract}


\end{frontmatter}

\section{Introduction}
Due to innovative developments in biotechnology, including tissue engineering and cell engineering, studies on active fluids composed of biological entities have recently drawn increasing attention from physical and mathematical scientists.
The pattern formations and fluid flows that are spontaneously generated by the internal motions of active entities, which interact with the local environment, are studied from a physical point of view at both the microscopic and macroscopic levels.
A suspension of chemotactic bacteria, e.g., \textit{E. Coli}, is a typical example of the active fluid, in which the bacteria create the collective motions, and the macroscopic patterns are created spontaneously via interactions with chemical cues, whose local concentrations also vary.\cite{art:75A,art:91BB,art:95BB,book:03B}

The macroscopic transport phenomena created by the collective motion of chemotactic bacteria can be modeled using the coupled reaction-diffusion equations for nutrients (which are consumed by bacteria), chemoattractants (which are secreted by bacteria), and bacterial density.
The basic idea of utilizing the reaction-diffusion equation to describe the collective motion of chemotactic bacteria was first introduced by Keller and Segel\cite{art:71KS,art:71KS2}, and both mathematical and physical studies on the Keller--Segel model have accumulated.
Although the microscopic basis of the Keller-Segel model was more recently laid by the asymptotic analysis of the kinetic chemotaxis model,\cite{art:08TMPA} this type of modeling is originally based on a phenomenological point of view.
The macroscopic transport phenomena are consequences of the migrations of bacteria, molecular diffusion of chemical cues (i.e., nutrients and chemoattractants), and their mutual interactions. \cite{art:74AT,art:72MK,art:83BSB,art:09KJTW}
Thus, the transport phenomena of chemotactic bacteria set an important multiscale problem to be resolved from physical and mathematical points of view.  
The mesoscopic model to describe the connection between the microscopic dynamics of bacterium and the macroscopic transports of chemical cues and bacteria population density takes on major significance.

The kinetic approach to describing the collective motion of chemotactic bacteria was first introduced by Alt \cite{art:80A} and then further developed \cite{art:88ODA}.
In the collective motions of the bacteria, each bacterium repeats a simple run-and-tumble motion, where they run ballistically in some duration and subsequently change their directions randomly.
In kinetic modeling, the run-and-tumble motion of bacterium is assumed to be a stochastic process, and the time evolution of the density of bacteria with a velocity of ${\bm v}$, $f(t,{\bm x},{\bm v})$ is described by a variant of the Boltzmann transport equation for gases (although the present kinetic transport equation does not involve the quadratic operator for collisions).\cite{book:07S,book:07P}
The transition (or scattering) kernel in the kinetic chemotaxis equation involves a model response function that stochastically determines the run durations of each bacterium according to the environment experienced by the bacteria along their pathways.
The response function takes a major role to reproduce the chemotactic motions of each bacterium and their collective behaviors. 
Thus, the kinetic approach is a promising candidate to investigate the connection between the microscopic dynamics of bacteria and macroscopic transport phenomena.

Recently, the studies on the kinetic chemotaxis have been actively performed by mathematical scientists.
Several researchers have succeeded in deriving the macroscopic continuum equations for chemotactic bacteria from the kinetic transport equation, and the connection between the macroscopic continuum description and the mesoscopic kinetic description has gradually been revealed.\cite{art:00HO,art:02OH,art:04CMPS,art:05EO,art:05DS,art:08SS,art:10SCBBSP,art:12LGS,art:13JV,art:15AEV,art:15X}
A comprehensive mathematical study on the traveling wave of the kinetic chemotaxis model is also given by Calvez.\cite{art:16C}
The kinetic chemotaxis model has also been utilized in the analysis of the experimental results. 
Saragosti \textit{et al} described the largely biased motions of bacteria toward regions with a high concentration of chemical cues, and they proposed a kinetic model based on experimental observations.\cite{art:11SCBPBS}

The numerical simulations of the kinetic chemotaxis model also takes on a significance in understanding the multiscale mechanism in the collective motions of chemotactic bacteria or analyzing the problems occurring in practical engineering and biological systems.
In the simulation methodologies for solving the kinetic chemotaxis models, the difficulty arises in the treatment of the response function in the scattering kernel.
It should be noted that the response function depends not only on the instantaneous spatial distribution of chemical cues but on the temporal variations along the running directions of each bacterium.
Recently, a Cartesian- mesh-based numerical method to accurately solve the kinetic chemotaxis equation was developed by Yang and Filbet and applied to various one- and two-dimensional problems.\cite{art:14YF}
In the method, an elaborate numerical algorithm is employed to treat the scattering kernel of the kinetic transport equation.

In the present study, a Monte Carlo simulation method for chemotactic bacteria in a three-dimensional system is newly developed on the basis of the kinetic chemotaxis equation used in Ref. \cite{art:11SCBPBS}, and the Monte Carlo method is applied to the traveling population wave in a microchannel.\cite{art:96G,art:03PWYLDSSA,art:06FMT,art:06SZLJL}
Because the Monte Carlo simulation employs a particle-based method, the treatment of the response function, which may depend on the memory of bacterium along its pathway\cite{art:96G,art:84AH,art:03MBBO,art:04G,art:05CG,art:15PTV}, is simplified.
Some numerical studies toward this direction were also put forward by Rousset and Samaey.\cite{art:13RS,art:13RS2}
The microscopic dynamics of bacterium and its relation to the macroscopic transports of chemical cues are investigated in detail.
The simulation results are also compared to the asymptotic solution obtained from the kinetic chemotaxis model in order to validate the accuracy of the Monte Carlo method in the asymptotic behaviors.
Thus, the purpose of the present study is to propose a new Monte Carlo simulation method and to demonstrate the validity and utility of the method numerically.
Incidentally, other Monte Carlo simulations, e.g., the Brownian dynamics simulations\cite{art:13FFCE,art:14RFE} and the velocity-jump simulation involving a memory kernel\cite{art:14MVSCM}, for the kinetic chemotaxis are also proposed. 
In the present Monte Carlo method, the velocity jump process described by the kinetic chemotaxis equation involving a response function of the tumbling rate and a persistence of the reorientation angle is utilized.

In Sec. II, the basic model, i.e., the kinetic chemotaxis model, and its non-dimensional form are described.
In Sec. III, the methodology of the present Monte Carlo simulation is explained in detail.
In Sec. IV, the numerical results of the Monte Carlo simulations are presented for the macroscopic transports, the microscopic dynamics, and the effects of variations in the parameters of the response function. Special focus is placed on the relation between the microscopic dynamics of bacterium and the macroscopic transports of the chemical cues and bacteria population density.
In Sec. V, the results of Monte Carlo simulations are compared to the asymptotic solution of the kinetic chemotaxis model in the continuum limit to examine the accuracy of the Monte Carlo method in the asymptotic behaviors.
Finally, concluding remarks and an outlook are given in Sec. VI.

\section{Basic Model} 
The basic model is described in the reference \cite{art:11SCBPBS}.
In the modeling, the macroscopic transport of chemical cues, i.e., the nutrients consumed by bacteria and the chemoattractants secreted by bacteria, is described by continuum reaction-diffusion equations, whereas the run-and-tumble motions of bacteria are described by a kinetic equation.
Because a small bacterium, e.g., \textit{E. coli}, is assumed, the cells are not able to choose directly the preferential direction of motion toward the region of a high concentration of chemical cues by measuring the head-to-tail gradient of the chemical cues.
Instead, the cells detect the preferential direction by sensing the temporal variation of chemical cues experienced along their pathways.
This sensing strategy of bacteria is accounted for in the response function in the kinetic equation.

\subsection{Basic equations}
In the following, the basic kinetic equations are explained briefly.
$f(t,{\bm x}, {\bm v})$ represents the density of bacteria with a velocity $\bm v$ at a time $t$ and a position $\bm x$, and $N(t,{\bm x})$ and $S(t,{\bm x})$ represent the concentration fields of nutrient and chemoattractant, respectively, at a time $t$.
These quantities are described by the following equations:
\begin{equation}
	\frac{\partial S}{\partial t}=
	D_S\frac{\partial^2 S}{\partial x_\alpha^2} - a S +b \int_{\bm v'\in V} f(t,{\bm x},{\bm v}')d\,{\bm v}',\label{trans_S}
\end{equation}
\begin{equation}
	\frac{\partial N}{\partial t}=
	D_N\frac{\partial^2 N}{\partial x_\alpha^2} - c N \int_{\bm v'\in V} f(t,{\bm x},{\bm v}')d\,{\bm v}',\label{trans_N}
\end{equation}
and
\begin{align}\label{trans_f}
	\frac{\partial f}{\partial t}+v_\alpha\frac{\partial f}{\partial x_\alpha}=&
	\int_{\bm v' \in V} T({\bm v},{\bm v}')f(t,{\bm x},{\bm v}')d\,{\bm v}'
	-\int_{\bm v' \in V} T({\bm v}',{\bm v})f(t,{\bm x},{\bm v})d\,{\bm v}' \nonumber \\
	&+rf(t,{\bm x},{\bm v}). 
\end{align}
Here, $D_S$ and $D_N$ are the diffusion coefficients for the chemoattractant and nutrient, respectively, $a$ is the degradation rate of chemoattractant, $b$ is the production rate of chemoattractant by a bacterium, and $cN$ is the consumption rate of nutrient by a bacterium.
Hereafter, the subscripts $\alpha$, $\beta$, and $\gamma$ represents the index in Cartesian coordinates, i.e., $\{\alpha,\beta,\gamma\}=\{x,y,z\}$.
Equation (\ref{trans_f}) is a variant of the Boltzmann equation for gases; in Eq. (\ref{trans_f}), the transition (scattering) kernel $T({\bm v},{\bm v}')$ stands for the tumbling event of a bacterium; during this event, the bacterium changes from velocity ${\bm v}'$ to ${\bm v}$.
The velocity space $V$ is bounded and symmetric.
In this study, we solely consider the case for bacteria with a preferential velocity $V_0$.
Thus, the integral domain $V$ represents the surface of sphere of a radius $V_0$, i.e., $V=\{{\bm v}|\sqrt{{\bm v}^2}=V_0\}$.
The last term in Eq. (\ref{trans_f}) represents cell division, where $r$ is the division rate of a bacterium ($r=\ln2/\tau_2$, where $\tau_2$ is the mean doubling time).

The transition kernel $T({\bm v},{\bm v}')$ is assumed to be split into two contributions; one is the tumbling rate $\lambda({\bm v}')$, and the other is the reorientation effect during tumbles $K({\bm v},{\bm v}')$.
\begin{equation}\label{split_T}
	T({\bm v},{\bm v}')=\lambda({\bm v}') K({\bm v},{\bm v}'),
\end{equation}
with the condition 
\begin{equation}\label{cond_K}
	\int_{\bm v \in V}K({\bm v},{\bm v}')d{\bm v}=1.
\end{equation}

For the tumbling rate $\lambda({\bm v})$, we assume that the bacteria are sensitive to the temporal variations of chemical cues along their pathways via logarithmic sensing mechanics \cite{art:72MK,art:83BSB,art:09KJTW} and that the contributions of each chemical cue are independent and additive.
Then, the tumbling rate $\lambda({\bm v})$ can be written as
\begin{subequations}
	\begin{align}
		\lambda({\bm v}')
		&=\frac{1}{2}\left(\lambda_N({\bm v}')+\lambda_S({\bm v}')\right) \nonumber \\
		&=\frac{1}{2}\left[\psi_N\left(\left.\frac{D\log N}{Dt}\right|_{\bm v'}\right) + \psi_S\left(\left.\frac{D\log S}{Dt}\right|_{\bm v'}\right)\right],
	\end{align}
\end{subequations}
where the $\left.\frac{D}{Dt}\right|_{\bm v'}$ is the material derivative along the trajectory with a velocity ${\bm v}'$.
The response functions $\psi_N$ and $\psi_S$ are both positive and decreasing because bacteria are less likely to tumble (thus perform longer runs) when the chemical cues increase.
In the present model, the following analytic function is chosen as the response function $\psi$;
\begin{equation}\label{eq_psi}
	\psi(X)=\psi_0-\chi\tanh\left(\frac{X}{\delta}\right),
\end{equation}
where $\psi_0$ is the basal mean tumbling frequency, $\chi$ is the modulation amplitude, and $\delta^{-1}$ is the characteristic time, which represents the stiffness of the response function.

$K({\bm v},{\bm v}')$ accounts for persistence in the successive run trajectories after the tumbles. \cite{art:07L}
When we consider the uniform scattering kernel, $K$ is a constant, i.e., $K=\frac{1}{4\pi V_0^2}$, and $T({\bm v},{\bm v}')$ is proportional to the tumbling rate $\lambda({\bm v}')$.
Persistence in the successive runs, ${\bm v}'\,\rightarrow\,{\bm v}$, can be described using the reorientation angle $\theta$, i.e.,
\begin{equation}\label{eq_theta}
	\cos\theta=\frac{{\bm v}\cdot{\bm v}'}{V_0^2},
\end{equation}
as
\begin{equation}
	K({\bm v},{\bm v}')\propto G\left(-\frac{1-\cos\theta}{\sigma^2}\right),
\end{equation}
where $\sigma$ is the standard deviation of reorientation angle $\theta$, i.e., $\sigma=\sqrt{<\theta^2>}$, and $G(X)$ is an increasing function, and its proportionality constant is determined using the normalization condition Eq. (\ref{cond_K}). 
For $G(X)=\exp(X)$, one can write the Von Mises distribution,
\begin{equation}
	K({\bm v},{\bm v}')=\frac{\exp\left(-\frac{1-\cos\theta}{\sigma^2}\right)}{2\pi V_0^2\sigma^2\left(1-e^{-\frac{2}{\sigma^2}}\right)}.
\end{equation}
Experiments also demonstrated that the tumbling frequency is strongly correlated with the reorientation angle.\cite{ art:11SCBPBS}
Thus, one can assume a linear correlation between the standard deviation of reorientation angle $\sigma$ and the tumbling frequency $\lambda({\bm v}')$ as 
\begin{equation}\label{eq_sigma}
	\sigma=\sigma_1+\sigma_2\lambda({\bm v}').
\end{equation}

Note that the rotational diffusivity and long-term memory of the bacteria are not involved in the present kinetic chemotaxis model.
\subsection{Non-dimensionalization}\label{subsec_non_dim} 
We introduce the non-dimensional time $\hat t$, space $\hat {\bm x}$, and velocity $\hat {\bm e}$, which are defined as
\begin{gather}
	\hat {\bm x}={\bm x}/L_0,\quad \hat {\bm e}={\bm v}/V_0,\quad \hat t=t/t_0\quad \left(t_0=L_0/V_0\right).
\end{gather}
We introduce the characteristic length $L_0$, which may affect the geometry of problem.
The bacterial density $f(t,{\bm x}, {\bm v})$ is also non-dimensionalized as
\begin{equation}
	\hat f(\hat t,\hat {\bm x}, \hat {\bm e})=f(t,{\bm x}, {\bm v})/(\rho_0/V_0^3).
\end{equation}

Using the non-dimensional quantities, Eqs. (\ref{trans_S}) -- (\ref{trans_f}) are written as follows.
\begin{equation}
	\frac{\partial \hat S}{\partial \hat t}=
	\hat D_S\frac{\partial^2 \hat S}{\partial \hat x_\alpha^2} - \hat a \hat S +\hat b \int_{{\rm all}\,\,\hat {\bm e}'} \hat f(\hat t,\hat {\bm x},\hat {\bm e}')d\,\Omega(\hat {\bm e}'),\label{neq_S}
\end{equation}
\begin{equation}\label{neq_N}
	\frac{\partial \hat N}{\partial \hat t}=
	\hat D_N\frac{\partial^2 \hat N}{\partial \hat x_\alpha^2} - \hat c \hat N\int_{{\rm all}\,\,\hat {\bm e}'} \hat f(\hat t,\hat {\bm x},\hat {\bm e}')d\,\Omega(\hat{\bm e}'),
\end{equation}
where $\hat D_{N,S}=D_{N,S}/(L_0^2/t_0)$, $\hat a=t_0 a$, $\hat b=1$, and $\hat c=\rho_0 t_0 c$.
The concentration of chemoattractant $S$ is scaled by the reference quantity $S_0=\rho_0 t_0 b$ and that of the nutrient $N$ is scaled by an arbitrary reference quantity $N_0$.
Note that the effect of the initial average density $\rho_0$ of bacteria is involved only via the non-dimensional parameter $\hat c$.

The kinetic chemotaxis equation of bacterial density $f(t,{\bm x},{\bm v})$ is written in non-dimensional form as
\begin{align}
	\frac{\partial \hat f}{\partial \hat t}+\hat e_\alpha\frac{\partial \hat f}{\partial \hat x_\alpha}=&
	\int_{{\rm all}\,\hat {\bm e}'} \hat \lambda(\hat {\bm e}')\hat K(\hat {\bm e},\hat {\bm e}')\hat f(\hat t,\hat {\bm x},\hat {\bm e}')d\,\Omega(\hat {\bm e}')
	-\hat \lambda(\hat {\bm e})\hat f(\hat t,\hat {\bm x},\hat {\bm e})
	+\hat r \hat f(\hat t,\hat {\bm x},\hat {\bm e}), \label{neq_f}
\end{align}
where $\hat\lambda=(L_0/V_0)\lambda$, $\hat K=V_0^3 K$, and $\hat r=(L_0/V_0) r$.
We also write $\hat \lambda(\hat{\bm e}')$ as $\hat \lambda(\hat{\bm e}')=\hat \psi_0\hat \Psi(\hat{\bm e}')$, where $\hat \psi_0=(L_0/V_0)\psi_0$, and the modulation of the tumble frequency $\Psi(\hat{\bm e}')$ is written as
\begin{equation}\label{neq_Psi}
	\hat \Psi(\hat {\bm e}')
	=\frac{1}{2}\left[\hat \psi_N\left(\left.\frac{D\log \hat N}{D \hat t}\right|_{\hat {\bm e'}}\right) + \hat \psi_S\left(\left.\frac{D\log \hat S}{D \hat t}\right|_{\hat {\bm e'}}\right)\right],
\end{equation}
with
\begin{equation}\label{neq_psi2}
	\hat \psi_{S,N}(X)=1-\hat \chi_{S,N} \tanh\left(\frac{\hat X}{\hat \delta}\right).
\end{equation}
Here $\hat \chi=\chi/\psi_0$ and $\hat \delta=(L_0/V_0)\delta$.
Note that $\hat \psi_0^{-1}(=(V_0\psi_0^{-1})/L_0)$ corresponds to the Knudsen number in the rarefied gas dynamics because the product of the bacterial velocity $V_0$ and the inverse of the mean tumble frequency $\psi_0^{-1}$, which is the mean free time, represents the mean free path.
$\hat K(\hat{\bm e},\hat{\bm e}')$ is written as
\begin{equation}\label{neq_K}
	\hat K(\hat{\bm e},\hat{\bm e}')=\frac{\exp\left(-\frac{1-\hat{\bm e}\cdot\hat{\bm e}'}{\sigma^2}\right)}{2\pi \sigma^2\left(1-e^{-\frac{2}{\sigma^2}}\right)}.
\end{equation}
$\hat K(\hat{\bm e},\hat{\bm e}')$ also satisfies $\int_{{\rm all}\,\hat {\bm e}}\hat K(\hat{\bm e},\hat{\bm e}')d\Omega(\hat {\bm e})=1$.

The standard deviation of reorientation angle $\sigma$ can be written as 
\begin{equation}\label{eq_hsigma}
	\sigma=\sigma_1+\sigma_2\hat \Psi(\hat{\bm e}'),
\end{equation}
where the constant values of $\sigma_1$ and $\sigma_2$ are determined from the maximum and minimum values of the standard deviation of reorientation angle, $\sigma_{\rm Max}$ and $\sigma_{\rm min}$, as
\begin{subequations}\label{det_sigma}
	\begin{align}
		\sigma_{\rm Max}=\sigma_1+\sigma_2\hat \Psi_{\rm Max},\\
		\sigma_{\rm min}=\sigma_1+\sigma_2\hat \Psi_{\rm min},
	\end{align}
\end{subequations}
where $\hat \Psi_{\rm Max}$ and $\hat \Psi_{\rm min}$ are the maximum and minimum values of Eq. (\ref{neq_Psi}), respectively, and are set to $\hat \Psi_{\rm Max}=1.4$ and $\hat \Psi_{\rm min}=0.6$ for Eqs. (\ref{neq_Psi}) and (\ref{neq_psi2}); these values are estimated from experimental results \cite{art:11SCBPBS}.

\section{Simulation Method}
\begin{figure}[htbp]
	\centering
	\includegraphics[scale=1]{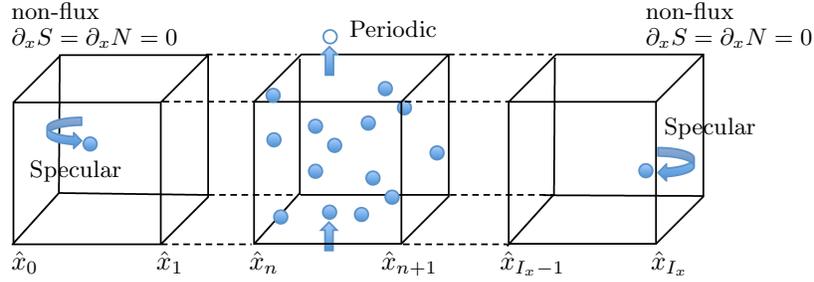}
	\caption{
		The lattice mesh system and the simulation particles in the lattice site. The run-and-tumble motions of particles are calculated by the Monte Carlo method while the concentrations of chemical cues are calculated by the finite volume method. The specular reflection condition for the particles and the non-flux condition for chemical cues are considered at $\hat x=\hat x_0$ and $\hat x_{I_x}$ while the periodic conditions are considered for both particles and chemical cues in $y$ and $z$ directions.
	}\label{pic_geom}
\end{figure}

In the present paper, we consider the one-dimensional macroscopic transports of chemical cues, i.e., $\partial S/\partial y=\partial S/\partial z=0$ and $\partial N/\partial y=\partial N/\partial z=0$, while the three-dimensional motions of bacteria are calculated by the Monte Carlo method as described below.
The spatial domain is discretized into a uniform lattice mesh system with $I_x\times I_y \times I_z$ cubes of a constant side length $\Delta \hat x$.
Here, $I_\alpha$ is the number of mesh intervals in the $\alpha$ direction of Cartesian coordinates.
The macroscopic transports of chemical cues are calculated on this lattice mesh system.
Because we consider the one-dimensional transports of chemical cues along $x$ coordinate, we can set $I_y=I_z=1$. 
The lattice mesh nodes are expressed as $\hat x_i\,(i\Delta \hat x)\,(i=0,1,\cdots,I_x)$ and the centers of each lattice are expressed as $\hat x_{i+\frac{1}{2}}\,(i=0,1,\cdots,I_x-1)$. See also Fig. \ref{pic_geom}.

Equations (\ref{neq_S}) and (\ref{neq_N}) are calculated with using a finite volume scheme,
\begin{equation}\label{eq_diffS}
	\frac{\hat S_i^{n+1}-\hat S_i^{n}}{\Delta \hat t}=
	\frac{\hat D_S}{\Delta \hat x}\left[
		\left(\frac{\partial \hat S}{\partial \hat x}\right)^{n}_{i+1}-
	\left(\frac{\partial \hat S}{\partial \hat x}\right)^{n}_{i}\right]
	-\hat a \hat S^{n+1}_i+\hat b\hat \rho^{n+1}_i,
\end{equation}
\begin{equation}\label{eq_diffN}
	\frac{\hat N_i^{n+1}-\hat N_i^{n}}{\Delta \hat t}=
	\frac{\hat D_N}{\Delta \hat x}\left[
		\left(\frac{\partial \hat N}{\partial \hat x}\right)^{n}_{i+1}-
	\left(\frac{\partial \hat N}{\partial \hat x}\right)^{n}_{i}\right]
	-\hat c\hat N^{n+1}_i \hat \rho^{n+1}_i,
\end{equation}
for $i=0,1,\cdots,I_x-1$, where the superscript $n$ represents the time step number, the subscript $i$ represents the $i$th lattice site, i.e., $\hat F^n_{i}=\hat F(n\Delta \hat t,\hat x_{i+\frac{1}{2}})$ ($F=\hat S$ or $\hat N$), and $(\partial F/\partial \hat x)^n_i$ is defined as
\begin{equation}\label{eq_gradF}
	\left(\frac{\partial F}{\partial \hat x}\right)^n_i=\frac{F^n_{i}-F^n_{i-1}}{\Delta \hat x},
\end{equation}
for $i=1,\cdots,I_x-1$. $(\partial F/\partial \hat x)_0$ and $(\partial F/\partial \hat x)_{I_x}$ are determined by the boundary conditions.
The macroscopic density of bacteria in the $i$th lattice site $\hat \rho_i$ is defined as
\begin{equation}\label{def_rho}
	\hat \rho_i^n=\frac{1}{\Delta \hat x^3}\int_{i{\rm th\, site}} \int_{{\rm all}\,\,\hat {\bm e}'}\hat f(n\Delta \hat t,\hat {\bm x},\hat {\bm e}')d\,\Omega(\hat {\bm e}')d\,{\hat {\bm x}},
\end{equation}
and is calculated from the number of simulation particles used in the Monte Carlo simulation as described in the next paragraph.
Note that the mesh interval $\Delta \hat x$ must be sufficiently small to resolve the macroscopic transports of chemical cues and the time-step size $\Delta t$ must be smaller than the diffusion time, i.e., $\Delta \hat t < 0.5\Delta \hat x^2/\hat D_{S,N}$.

The three-dimensional motions of bacteria are calculated in each lattice site by the Monte Carlo method.
We use a constant and uniform weight $w_0$ for a single simulation particle; this weight represents the number of bacteria corresponding to each simulation particle.
That is, the number of bacteria in the $i$th lattice site is written as
\begin{equation}\label{eq_w0}
	(L_0\Delta \hat x)^3 \rho_0\hat \rho_i=
	w_0{\mu_i},
\end{equation}
where $\mu_i$ is the number of simulation particles contained in the $i$th lattice site.
Thus, the macroscopic density at the $i$th lattice site $\hat \rho_i$ is calculated from the number of simulation particles $\mu_i$ via Eq. (\ref{eq_w0}). 
Note that the initial total number of simulation particles $M_0$ and the weight $w_0$ are not related to the physical results but to the accuracy of the Monte Carlo simulation.
The effect of the initial mean density of bacteria $\rho_0$ is involved via the non-dimensional parameter $\hat c$.

The Monte Carlo simulation is conducted using the following steps.
Hereafter, the position and velocity of the $l$th particle are expressed as $\hat {\bm r}_{(l)}$ and $\hat {\bm e}_{(l)}$, respectively.

\begin{enumerate}
	\setcounter{enumi}{-1}
\item At $\hat t=0$, the simulation particles are distributed according to the initial density $\hat f^0_i(\hat {\bm e})$ ($i=0,\cdots,I_x-1$).
		The number of simulation particles in the $i$th lattice site, $\mu_i^0$ is determined by Eq. (\ref{eq_w0}).
		In each lattice site, simulation particles are distributed uniformly at random positions and their velocities $\hat {\bm e}$ are determined by the probability density $\hat f_i^0(\hat {\bm e})/\hat \rho_i^0$.
	\item 
		Particles move with their velocities for a duration $\Delta \hat t$:
		\begin{equation}
			\hat {\bm r}_{(l)}^{n+1}=\hat {\bm r}_{(l)}^n+\hat {\bm e}_{(l)}^n\Delta \hat t\quad (l=1,\cdots,M^n),
		\end{equation}
		where $M^n$ is the total number of simulation particles at time step $n$.
		The particles that move beyond the boundaries are removed, and new ones are inserted according to the boundary conditions.

		The number of particles at each lattice site, $\mu_i^{n+1}$ ($i=0,\cdots,I_x-1$) is also counted.
	\item
		At each lattice site, the concentrations of chemical cues $\hat S^{n+1}_i$ and $\hat N^{n+1}_i$ ($i=0,\cdots,I_x-1$) are calculated by Eqs. (\ref{eq_diffS}) and (\ref{eq_diffN}) with Eq. (\ref{eq_w0}). 
	\item
		The tumbling of each particle is calculated using the scattering kernel in Eq. (\ref{neq_f}).
		The tumbling of the $l$th particle may occur with a probability $(\hat \psi_0 \Delta \hat t)\hat \Psi(\hat {\bm e}^n_{(l)})$.
	
		$\hat \Psi(\hat {\bm e}^n_{(l)})$ is calculated from the temporal variations of chemical cues along the pathway of the $l$th particle, $\hat {\bm r}^n_{(l)}\rightarrow \hat {\bm r}^{n+1}_{(l)}$, in the successive time steps.
		We write the chemical cues experienced by the $l$th particle at the time step $n$ as $\hat S_{(l)}^n$ and $\hat N_{(l)}^n$.
		Then, 
		\begin{equation}\label{eq_Psil}
			\hat \Psi(\hat {\bm e}^n_{(l)})=\frac{1}{2}\left({\psi_S}_{(l)}+{\psi_N}_{(l)}\right),
		\end{equation}
		with
		\begin{subequations}
			\begin{align}
				{\psi_S}_{(l)}&=1-\hat \chi_S\tanh\left(\frac{\log \hat S_{(l)}^{n+1}-\log \hat S_{(l)}^n}{\hat \delta \Delta \hat t}\right),\\
				{\psi_N}_{(l)}&=1-\hat \chi_N\tanh\left(\frac{\log \hat N_{(l)}^{n+1}-\log \hat N_{(l)}^n}{\hat \delta \Delta \hat t}\right).
			\end{align}
		\end{subequations}
		$\hat S_{(l)}^n$ and $\hat N_{(l)}^n$ are the local concentration of chemical cues at the position of the $l$th particle, $\hat {\bm r}^n_{(l)}$, and are calculated by the interpolation using the concentrations of the chemical cues at the lattice site in which the particle in involved and the linear gradients between the neighboring lattice sites. 
		Thus, when the $l$th particle is involved in the $i$th lattice site, $F_{(l)}^n$ ($F=\hat S$ or $\hat N$) is calculated as
		\begin{equation}\label{eq_linear}
			\hat F^n_{(l)}=\hat F^n_i + 
			\left(\frac{\partial F}{\partial \hat x}\right)^n_{i+\frac{1}{2}}
			(\hat {r_x}^n_{(l)}-{\hat x}_{i+\frac{1}{2}}), 
		\end{equation}
		where the linear gradient is calculated using the central difference between neighboring lattice sites as
		\begin{equation}\label{eq_gradF2}
			\left(\frac{\partial F}{\partial \hat x}\right)_{i+\frac{1}{2}}^n
			=\frac{1}{2}\left[
				\left(\frac{\partial F}{\partial \hat x}\right)^n_i+
				\left(\frac{\partial F}{\partial \hat x}\right)^n_{i+1}
			\right],
		\end{equation}
		where $(\partial F/\partial \hat x)_i$ is defined in Eq. (\ref{eq_gradF}).
		Using the linear interpolation Eq. (\ref{eq_linear}), a simulation particle that stays at the same lattice site after a single time step passes can sense the chemical gradients along its pathway.
		
		For the particle that is judged to tumble, say the $l'$th particle, a new run direction after the tumbling, $\hat {\bm e}^{n+1}_{(l')}$, is determined using the probability $\hat K(\hat {\bm e}^{n+1}_{(l')},\hat {\bm e}^{n}_{(l')})$ in Eq. (\ref{neq_K}).
		By using the stochastic variables $\theta$ and $\phi$ calculated as
		\begin{subequations}
			\begin{align}
				\cos\theta&=1+\sigma^2\log\left[
					e^{-\frac{2}{\sigma^2}}
					+(1-e^{-\frac{2}{\sigma^2}})U_1
				\right],\\
				\phi&=2\pi U_2,
			\end{align}
		\end{subequations}
		where $U_1$ and $U_2$ are uniform random numbers,
		the new run direction of the $l'$th particle is obtained by
		\begin{equation}
			\hat {\bm e}^{n+1}_{(l')}=
			\hat {\bm e}^n_{(l')}\cos\theta			
			+\hat {\bm e}^n_1 \sin\theta\cos\phi
			+\hat {\bm e}^n_2 \sin\theta\sin\phi,
		\end{equation}
		where $\hat {\bm e}^n_1$ and $\hat {\bm e}^n_2$ are the normal vectors to $\hat {\bm e}^n_{(l')}$ and are mutually perpendicular.
		For example, they can be written as
		\begin{subequations}
			\begin{align}
				\hat {\bm e}^n_1
				&=\left(
				\frac{e_y^n}{\sqrt{(e_x^n)^2+(e_y^n)^2}},
				\frac{-e_x^n}{\sqrt{(e_x^n)^2+(e_y^n)^2}},
				0
				\right),\\
				\hat {\bm e}^n_2
				&=\left(
				\frac{e_x^n e_z^n}{\sqrt{(e_x^n)^2+(e_y^n)^2}},
				\frac{e_y^n e_z^n}{\sqrt{(e_x^n)^2+(e_y^n)^2}},
				-\sqrt{(e_x^n)^2+(e_y^n)^2}
				\right).
			\end{align}
		\end{subequations}
	\item
		For all simulation particles, divisions are calculated with a uniform probability $\hat r\Delta t$.
		For a particle that is judged to undergo division, e.g., the $l$th particle, a new particle with the run 
		direction $\hat {\bm e}_{(l)}$ is created at a random position in the same lattice site.
		Note that the total number of simulation particles is updated as $M^{n+1}$.
	\item
		Return to step 1 with the obtained ${\hat {\bm r}}_{(l)}$, ${\hat {\bm e}}_{(l)}$, $\hat S_{(l)}$, and $\hat N_{(l)}$ ($l$=1,$\cdots$,$M$) at the new time step.
\end{enumerate}

\section{Results}

The simulation method described in the previous section is applied to the traveling population wave of bacteria in a microchannel. The parameter values used in the simulations, unless otherwise stated, are summarized in Tables \ref{t_reference} and \ref{t_parameters}. 
\begin{table}[htbp]
	\centering
	\caption{Reference quantities.}\label{t_reference}
	\begin{tabular}{cc}
		\hline\hline
		$L_0$  & 1 [mm]\\
		$V_0$  & 25 [$\mu$m/s]\\
		$t_0$  & 40 [s]\\
		$\rho_0$ & 5$\times 10^8$ [cell/mL]\\
		\hline\hline
	\end{tabular}
\end{table}

The model parameters are chosen to reproduce the experimental and numerical results obtained in Ref. \cite{art:11SCBPBS}. The numerical parameters $\sigma_1$ and $\sigma_2$ are determined from Eq. (\ref{det_sigma}) with $\sigma_{\rm Max}=1.5$ and $\sigma_{\rm min}=1.3$, which also correspond to the experimentally reported values\cite{art:11SCBPBS}. 

\begin{table}[htbp]
	\centering
	\caption{Parameter values used in simulations.}\label{t_parameters}
	\begin{tabular}{lcc}
		\hline\hline
		\multicolumn{3}{c}{Model parameters} \\
		\hline
		degradation rate of chemoattractant $a$&5$\times 10^{-3}$ [1/s]&(0.2)\\
		production rate of chemoattractant $b$&4$\times 10^{5}$ [1/cell/s]&\\
		consumption rate of nutrient $c$&$5\times 10^{-11}$ [mL/cell/s]&(1.0)\\
		diffusion coefficient of nutrient $D_N$&$8\times 10^{-6}$ [cm$^{2}$/s] &(0.032)\\
		diffusion coefficient of chemoattractant $D_S$&$8\times 10^{-6}$ [cm$^{2}$/s] &(0.032)\\
		mean tumble frequency $\psi_0$ & 3.0 [1/s]& (120) \\
		modulation of tumble freq. $\chi_N/\psi_0$  & 0.6 &\\
		modulation of tumble freq. $\chi_S/\psi_0$  &  0.2 &\\
		coefficient in Eq. (\ref{eq_sigma}) $\sigma_1$  & 0.85&\\
		coefficient in Eq. (\ref{eq_sigma}) $\sigma_2$  & 0.40&\\
		stiffness of the response functions $\delta^{-1}$  & 8 [s]&(0.2)\\
		mean doubling time $\tau_2=\ln2/r$ &1.15 [h] &(103.5) \\
		channel length $L$ &1.8 [cm] &(18)
		\\
		\multicolumn{3}{c}{Numerical parameters}\\
		\hline
		mesh interval $\Delta x$  & 25 [$\mu$m]&(0.025)\\
		time step size $\Delta t$  & 0.2 [s]&(0.005)\\
		total number of initial simulation particles &56640  &($w_0$=0.1)\\
		\hline\hline
		\multicolumn{3}{c}{The values inside the parentheses indicate the values for the non-dimensional forms.}
	\end{tabular}
\end{table}

The total number of simulation particles is ten times larger than the number of bacteria in the experimental system, i.e., the weight $w_0$=0.1, to obtain an accurate solution to Eqs. (\ref{neq_S})--(\ref{neq_f}) by reducing the noise that arises in the Monte Carlo method.

The one-dimensional lattice mesh system is used for the nutrient and chemoattractant fields, where the non-flux conditions are used at the channel edges $x$=0 and $L$, i.e., $(\partial \hat S/\partial \hat x)_0=(\partial \hat S/\partial \hat x)_{I_x}=0$ and $(\partial \hat N/\partial \hat x)_0=(\partial \hat N/\partial \hat x)_{I_x}=0$ in Eqs. (\ref{eq_diffS}) and (\ref{eq_diffN}). The boundary conditions for the simulation particles are periodic in the $y$ and $z$ directions and specular reflection at the channel edges $x$=0 and $L$. The initial conditions for $\hat S$ and $\hat N$ at $\hat t=0$ are uniform and given as $\hat S(\hat x)=0$ and $\hat N(\hat x)=1$. The simulation particles are exponentially distributed according to the initial density profile $\hat \rho(\hat x)$, i.e.,
\begin{equation}
	\hat \rho(\hat x)=\alpha\exp(-\beta\hat x),
\end{equation}
where $\alpha$ and $\beta$ satisfy the conditions $\int_0^{\hat L} \hat \rho(\hat x)d\hat x=1$ and $\int_0^{\hat w} \hat \rho(\hat x)d\hat x=0.99$ with $w=2\,[{\rm mm}]$.

\subsection{Macroscopic transport}

\begin{figure}[h]
	\centering
	\includegraphics[scale=1.0]{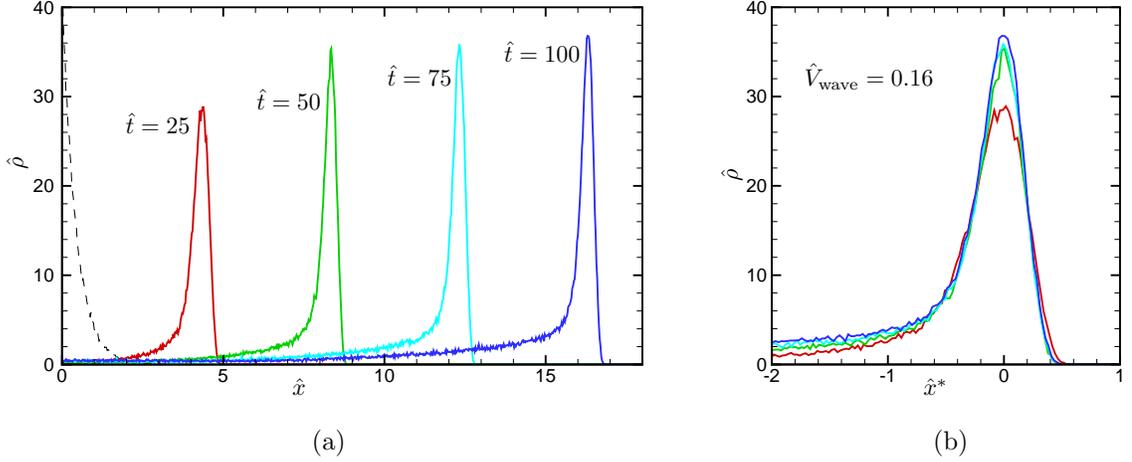}
	\caption{The traveling population wave of bacteria along the channel. (a) Snap shots of the density profiles of the bacteria and (b) the superposition of the snap shots in (a) using the relative coordinate $\hat x^*$, which is the position relative to the peak position of the density profile that moves with a constant traveling speed $\hat V_{\rm wave}$. The traveling speed is calculated as $\hat V_{\rm wave}$=0.16, i.e., $V_{\rm wave}$=4.0 $\mu$m/s. The dashed line in Fig. (a) shows the initial density profile of the bacteria. See also Supplemental Material for 3D motions of bacteria and time progress of bacterial density and nutrient concentration.
}\label{fig_twave_sposi}
\end{figure}

Figure \ref{fig_twave_sposi} shows the traveling population wave of bacteria along the channel. The population wave propagates with a constant velocity $\hat V_{\rm wave}$=0.16, i.e., $V_{\rm wave}$=4.0 [$\mu$m/s], which is close to the experimental measurement of $V_{\rm wave}$=4.1 [$\mu$m/s]. The wave profile changes only slightly after the initial transient period, $\hat t \gtrsim 50$; the density at the peak and rear side of the wave profile only slightly increases as time progresses, whereas the front-side profile of the wave is almost unchanged.

\begin{figure}[h]
	\centering
	\includegraphics[scale=1.0]{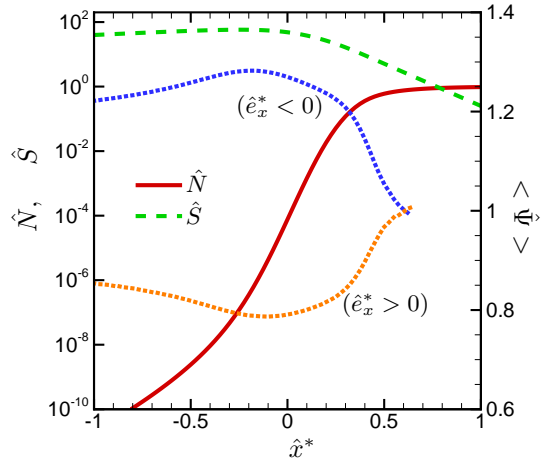}
	\caption{
		The spatial variations of the concentration of chemical cues, i.e., the nutrient $\hat N$ (solid line) and the chemoattractant $S$ (dashed line), and the mean tumbling rate of the bacteria $< \hat\Psi >$ (dotted lines) along the relative coordinate $\hat x^*$ are shown. (See also the caption of Fig. \ref{fig_twave_sposi}.) The upper blue and lower orange dotted lines show the mean tumbling frequency of bacteria with negative and positive velocities, respectively, relative to the traveling speed $\hat e^*$ (i.e., $\hat e_x^*=\hat e_x-\hat V_{\rm wave}$). The left and right vertical axes show the chemical cues and the mean tumbling frequency, respectively. 
	}\label{fig_psi_NS_delta008}
\end{figure}

The concentration profiles of chemical cues, i.e., the nutrient $N$ and the chemoattractant $S$, are coupled to the collective motion of bacteria via the consumption of nutrients and the production of chemoattractants by bacteria. The motions of bacteria are also affected by the concentration profiles of chemical cues.
Figure \ref{fig_psi_NS_delta008} shows the concentration profiles of chemical cues and the spatial variation of the mean tumbling rate $<\hat \Psi >$ along the relative coordinate $\hat x^*$, which is the position relative to the peak position that moves with a constant traveling speed $\hat V_{\rm wave}$. The local mean tumbling frequency is obtained by taking an ensemble average of the tumbling frequencies of each bacterium $\hat \Psi_{(l)}$ (see Eq. (\ref{eq_Psil})) within the lattice site. In Fig. \ref{fig_psi_NS_delta008}, every profile is also time-averaged over a time period $\hat t=$50 to $\hat t=$100.
Incidentally, in Fig. \ref{fig_psi_NS_delta008}, the mean tumbling rates $<\hat \Psi>$ are not shown in the region $\hat x^*\gtrsim 0.7$ because the bacteria are not in this region. 

The nutrient concentration exponentially increases as the relative coordinate $\hat x^*$ for $\hat x^*\lesssim 0.2$ and approaches unity for $\hat x^*\gtrsim 0.5$. The gradient takes a maximum around the position that coincides with the peak of the bacterial density, i.e., $\hat x^*\sim 0$, where the consumption of nutrient by bacteria concentrated. 
In contrast, the concentration of chemoattractant is not a monotonic function; it moderately increases as $\hat x^*$ at the rear side of the population wave, i.e., $\hat x^*\lesssim 0$ and takes a maximum around the peak of the density of bacteria, $\hat x^*\sim 0$; thus, the concentration profile of chemoattractant is similar to that for the bacterial density because the production of chemoattractant is proportional to the bacterial density. The chemoattractant concentration also exponentially decreases as $\hat x^*$ due to a simple diffusion process in the region ahead of the population wave.

The tumbling frequency of a bacterium depends on the material derivatives of chemical cues along the pathway of the bacterium. See Eq. (\ref{eq_Psil}). Because the profiles of chemical cues do not change much along the relative coordinate $\hat x^*$, the material derivative can be estimated as 
\begin{equation}\label{eq_xrc}
	\left. \frac{D}{D\hat t}\right |_{\hat e_x}\simeq (\hat e_x-\hat V_{\rm wave})\frac{\partial}{\partial \hat x^*}.
\end{equation}
Thus, in Fig. \ref{fig_psi_NS_delta008}, the modulation amplitudes of mean tumbling frequency for each of the positive and negative relative velocities are almost symmetric about the basal mean tumbling frequency $\hat \Psi=1$. The modulation amplitudes are magnified just behind the peak of the bacterial density, i.e., $\hat x^*\sim 0$, where the gradient of the nutrient takes a maximum and that of the chemoattractant is also non-negative; the modulation amplitudes decrease as $\hat x^*$ increases at the front side of the population wave $\hat x^*>0$, where both of the gradients of the chemical cues decrease.

\subsection{Microscopic dynamics}

\begin{figure}[h]
	\centering
	\includegraphics[scale=1.0]{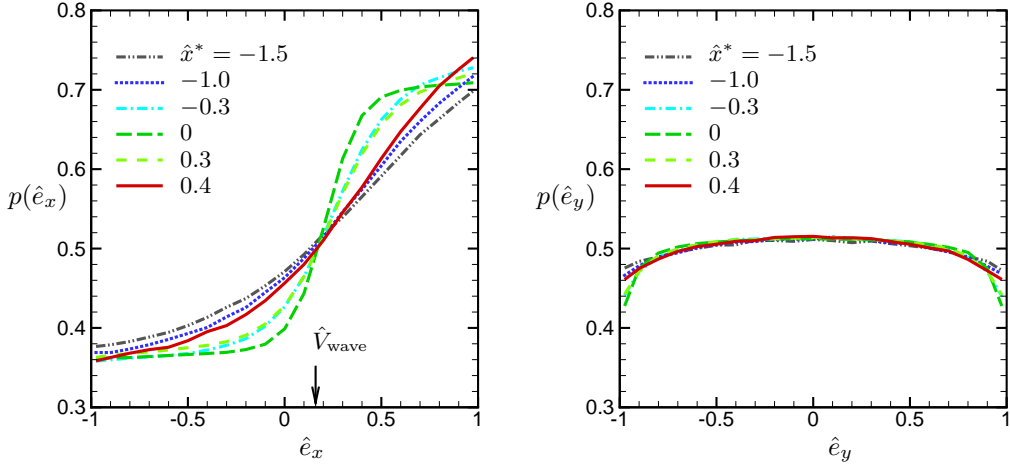}
	\caption{
		The probability density functions of the velocity of a bacterium $p(\hat e_\alpha)$ at different positions along the relative coordinate $\hat x^*$. (a) The PDFs for the longitudinal velocity and (b) for the lateral velocity. The downward arrow shows the traveling speed of the population wave.
	}\label{fig_pdfxy_delta008}
\end{figure}

The microscopic dynamics of the bacteria that form the traveling population wave are also investigated in terms of the probability density function (PDF) and the autocorrelation function (ACF) of the velocity of bacterium. 
Figure \ref{fig_pdfxy_delta008} shows the local PDFs $p(\hat e_\alpha)$ at different positions along the relative coordinate $\hat x^*$. The local PDFs are calculated by taking the time averages of the distribution of instantaneous velocities of the bacteria in each local lattice site for $\hat t$=[50,100]. 
Note that the simulations are performed with a weight $w_0$=0.01, to obtain accurate profiles of the PDFs.
The PDF of the longitudinal velocity, $p(\hat e_x)$, rapidly increases around the traveling speed of the population wave, $\hat e_x \sim \hat V_{\rm wave}$. 
The gradient around the traveling speed is larger in the region around the peak of the population wave, i.e., $\hat x^*$=0 and $\pm$0.3, than in the rear and front regions, $|\hat x^*|>0.3$.
In contrast, the PDF of the lateral velocity, $p(\hat e_y)$, is symmetric and rather flat except at large velocities, $\hat e_{y}\sim 1$. 
The spatial variation of the PDF of the lateral velocity is also small.
The steep gradient of the velocity distribution around the traveling speed in Fig. \ref{fig_pdfxy_delta008}(a) is related to the stiffness of the response function and the large spatial gradient of the nutrient around the peak.
A detailed discussion is given in \ref{appendix_over}.

\begin{figure}[h]
	\centering
	\includegraphics[scale=1.0]{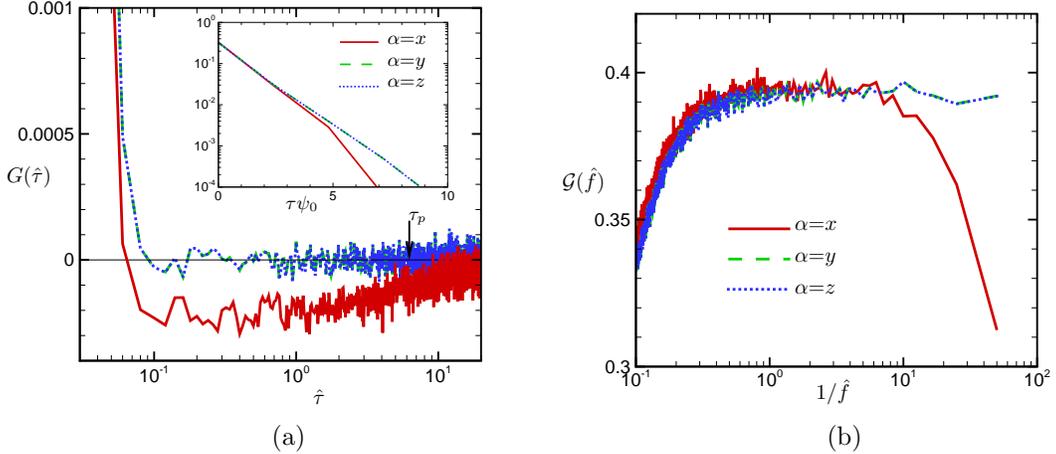}
	\caption{
		The autocorrelation function $G(\hat \tau)$ (a) and power spectrum ${\cal G}(\hat f)$ (b) of the deviation velocity, as defined in Eq. (\ref{eq_xi}), of bacteria within the traveling population wave. In the figures, $\hat \tau$ is the lag time, and $\hat f$ is the frequency, i.e., $1/\hat f$ is the period. The inset in (a) magnifies the behavior of the autocorrelation function in a short time period rescaled by the mean tumbling frequency $\psi_0$.
	}\label{fig_corr_wv_delta008}
\end{figure}

The ACF of the deviation velocity, which is defined in Eq. (\ref{eq_xi}), of the bacteria that form the traveling population wave are shown in Fig. \ref{fig_corr_wv_delta008}. 
The ACF $G(\hat \tau)$ is calculated from the trajectories of the velocity of each bacterium within a concentrated region, where the local density is not less than 10 \% of the peak density at $\hat t$=100, i.e.,
\begin{equation} 
	G(\hat \tau)=\overline{\left<{\xi_\alpha}_{(l)}(\hat t){\xi_\alpha}_{(l)}(\hat t-\hat \tau)\right>}, \quad(\alpha=x,y,z),\label{eq_Gt}
\end{equation}
where
\begin{equation}
	{\xi_\alpha}_{(l)}(\hat t)=\hat {e_\alpha}_{(l)}(\hat t)-\left<\hat {e_\alpha}_{(l)}(\hat t)\right>.\label{eq_xi}
\end{equation}
Here, $\left<\cdot\right>$ represents the ensemble average of the test particles, and $\overline{A(\hat t)}$ represents the time average of $A(\hat t)$ over $\hat t=[50,100]$.
The power spectrum ${\cal G}(\hat f)$ is calculated using the Fourier transform of $G(\hat \tau)$ as ${\cal G}(\hat f)=\int_0^{50} G(\hat \tau)\exp(-i2\pi \hat f \hat \tau)d\hat \tau$. 
Note that the inset in Fig. \ref{fig_corr_wv_delta008}(a) magnifies the behavior of the ACF in a short time period scaled by the mean tumbling frequency $\psi_0$.

The ACFs of the lateral deviation velocities, i.e., $\alpha$=$y$ and $z$, exponentially decrease on the short time scale, so that the autocorrelation of the lateral deviation velocities almost vanish after several tumbling events. 
These features of the lateral deviation velocity demonstrate the simple Poisson process of the tumbling events of bacteria for movements in lateral directions.

However, the ACF of the longitudinal deviation velocity exhibits considerably different behavior from that of the lateral deviation velocity. 
Notably, a negative autocorrelation arises for the longitudinal deviation velocity after a rapid decline on the short time scale due to the Poisson tumbling process. 
The amplitude of the negative correlation is small, but the negative correlation extends for a long time period ($\gg 1/\hat \psi_0$). 
The power spectrum of the longitudinal deviation velocity also shows a decline in the long time period.
Thus, the negative autocorrelation function and its power spectrum characterize the random motions of the bacteria trapped within the population wave. 
The time period of the negative autocorrelation, $\hat \tau_p$, can be estimated as the characteristic time in which the traveling population wave passes through the width of density profile $W_\rho$, estimated as $W_\rho\sim L_0$, i.e., $\hat \tau_p = 1/\hat V_{\rm wave}$. 
\subsection{Effect of variations in the parameters of the response function}
In the present kinetic model, the microscopic dynamics of each bacterium is coupled to the macroscopic transport of chemical cues via the response function defined in Eq. (\ref{eq_psi}), in which the sensitivity of the bacterium and the modulation amplitude of the tumbling frequency are characterized by the stiffness parameter $\delta^{-1}$ and the modulation parameters $\chi_N$ and $\chi_S$, respectively. 
The solutions obtained using the present kinetic model are significantly affected by those parameters. 
As the stiffness parameter $\delta^{-1}$ increases, the profile of the response function $\psi(X)$ becomes similar to a step function; thus, the tumbling rate of the bacterium switches as soon as the sign of the gradient of the chemical cue along the pathway changes. 
As the modulation parameter $\chi$ increases, the difference in the mean tumbling frequencies of the bacteria for the negative and positive run velocities becomes larger; thus, the biased motion of the bacterium toward the migration direction is enhanced.
In this subsection, the effects of changing the parameters of the response function are investigated. Monte Carlo simulations are performed for various values of the stiffness parameter, i.e., $\hat \delta^{-1}$=0.125, 0.15, 0.2, 0.25, 0.5, and 1.0, and the modulation parameter of nutrient, i.e., $\hat \chi_N$=0.5, 0.55, 0.6, 0.65, 0.7, and 0.8.

\begin{figure}[htbp]
	\centering
	\includegraphics*[scale=1.0]{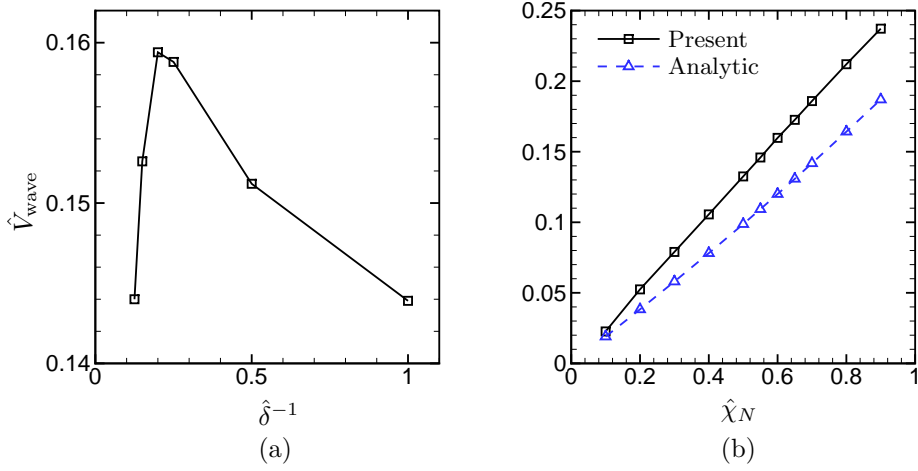}
	\caption{
		The traveling speed of the population wave vs. the stiffness parameter $\hat \delta^{-1}$ (a) and the modulation parameter $\hat \chi_N$ (b) of the response function.
		The squares $\square$ show the present results.
		Note that, for the present results, the modulation parameter is fixed as $\hat \chi_N$=0.6 in figure (a), the stiffness parameter is fixed as $\hat \delta^{-1}$=0.2 in figure (b), and the mean tumbling frequency is fixed as $\hat \psi_0$=120 in both figures. 
		The triangles $\triangle$ in figure (b) show those obtained by the analytical formula in Ref. \cite{art:11SCBPBS} for the stepwise response function (i.e., $\hat \delta^{-1}\rightarrow \infty$) in the continuum limit where the Knudsen number vanishes.
	}\label{fig_tvel}
\end{figure}

\begin{figure}[h]
	\centering
	\includegraphics[scale=1.0]{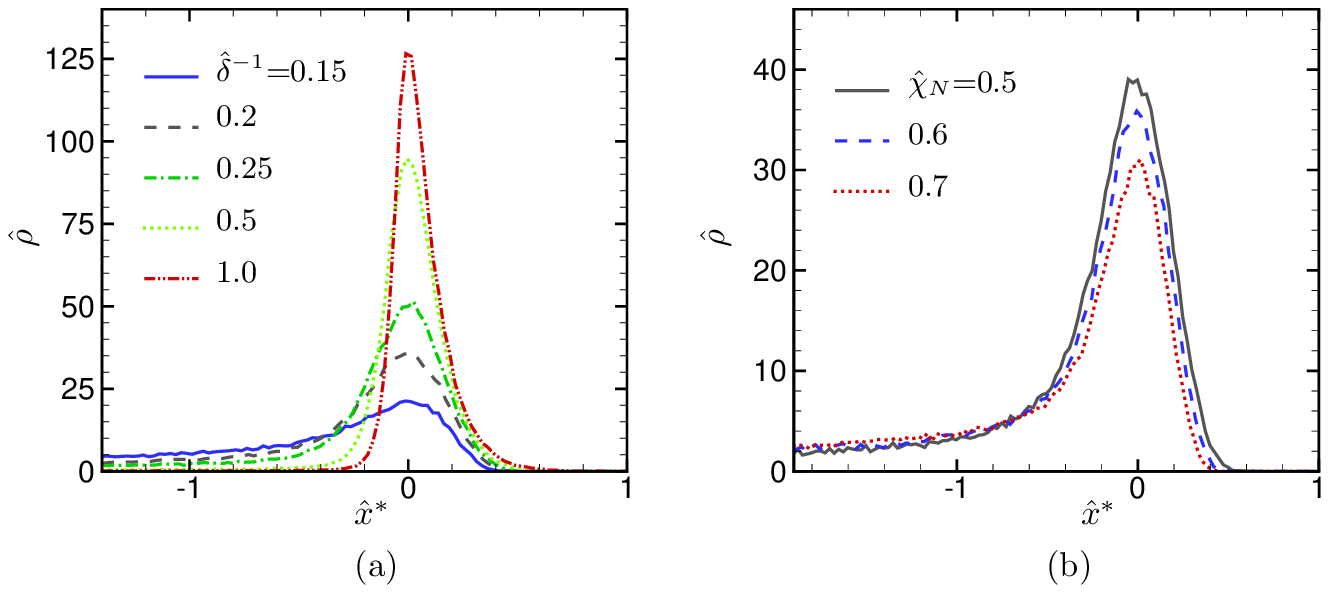}
	\caption{The superposition of the density profiles at $\hat t$=75 along the relative coordinate $\hat x^*"$ with variations in the stiffness parameter $\hat \delta^{-1}$ (a) and modulation parameter of the nutrient $\hat \chi_N$ (b).}\label{fig_sposi_deltas_chis} 
\end{figure}

Figures \ref{fig_tvel} and \ref{fig_sposi_deltas_chis} show the traveling speed and density profiles of population waves for different values of the stiffness parameters $\hat \delta^{-1}$ and the modulation parameter of nutrient $\hat \chi_N$, respectively.
The stiffness parameter of the response function $\hat \delta^{-1}$ does not substantially affect the traveling speed but significantly affects the symmetry of the wave profile. See Figs. \ref{fig_tvel}(a) and \ref{fig_sposi_deltas_chis}(a). 
As the stiffness increases, the width of the density profile becomes thinner, and the peak of the density increases. 
The symmetry of the density profile also changes. 
For small values of stiffness, the tail created behind the population wave extends to a broader range. 
However, for the large values of stiffness, i.e., $\hat \delta^{-1}$=0.5 and 1.0, the density declines very steeply behind the peak, so the front side of the population wave becomes broader than the rear side. 
The traveling speed of the population wave $\hat V_{\rm wave}$ shows a non-monotonic dependency of the stiffness. 
The traveling speed is maximized at $\hat \delta^{-1}$=0.2 in the present simulations.
The mechanism for the maximum arising in the traveling speed with variations in the stiffness parameter is not clearly known. 
However, it seems to be related to the relationship between the sensitivity of bacterium to the nutrient and the concentration profile of nutrient produced by the bacteria; as the stiffness parameter increases, the response of bacterium to the nutrient becomes more sensitive, but the width that the bacteria climb coherently along the gradient of nutrient become narrower.

In contrast, the modulation parameter for nutrient in the response function $\hat \chi_N$ does not substantially affect the profile of the population wave. The symmetry of the wave profile does not change, but the peak of the density decreases inversely to the length of the tail behind the population wave as the modulation parameter $\hat \chi_N$ increases. However, the traveling speed is linearly related to the modulation parameter: $\hat V_{\rm wave} \propto \hat\chi_N$.
This linear dependency is also observed in the results obtained by the analytical formula of the traveling speed for the stepwise response function in the continuum limit, which is obtained in Ref. \cite{art:11SCBPBS}.
The discrepancy between the results of the present Monte Carlo simulations and analytical formula arises because the present results are obtained for the finite values of the Knudsen number $\hat \psi_0^{-1}$ and stiffness parameter $\hat \delta^{-1}$ while, in Ref. \cite{art:11SCBPBS}, the limiting case for those parameters, i.e., $\hat \psi_0^{-1}\rightarrow 0$ and $\hat \delta^{-1}\rightarrow\infty$, is considered.
The convergence of the traveling speed obtained by the Monte Carlo method in that limiting case can be seen in Fig. \ref{fig_tvel_eps}.

\begin{figure}[h]
	\centering
	\includegraphics[scale=1.0]{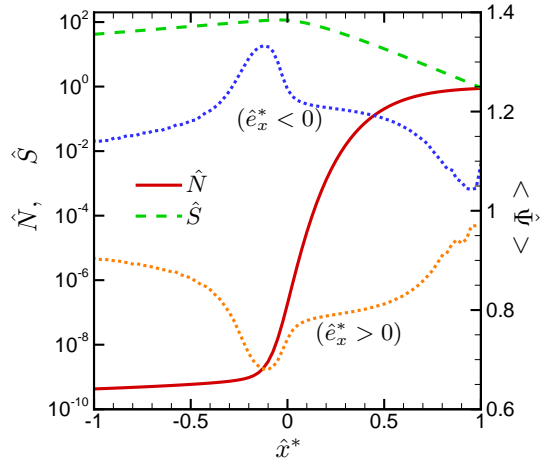}
	\caption{
		The spatial variations of the concentration of the nutrient $\hat N$ (solid line) and the chemoattractant $\hat S$ (dashed line) and that of the mean tumbling rate of bacteria $<\hat \Psi>$ (dotted lines) along the relative coordinate $\hat x^*$ for the stiffness parameter $\hat \delta^{-1}$=1.0. See also the caption in Fig. \ref{fig_psi_NS_delta008}
	}\label{fig_psi_NS_delta040}
\end{figure}

The profiles of the concentration of chemical cues, $\hat N$ and $\hat S$, and the spatial variation of mean tumbling frequency $<\hat \Psi>$, which are shown in Fig. \ref{fig_psi_NS_delta008}, are not substantially affected by changes in the modulation parameter $\hat \chi_N$. 
The stiffness parameter $\hat \delta^{-1}$ also does not substantially affect the profiles in the range of $\hat \delta^{-1}$=0.125 to 0.25.
However, for large stiffness parameter, i.e., $\hat \delta^{-1}=$0.5 and 1.0, the profiles are significantly changed from the profile shown in Fig. \ref{fig_psi_NS_delta008}. 
Figure \ref{fig_psi_NS_delta040} shows the spatial variations of the chemical cues and the mean tumbling frequency along the relative coordinate $\hat x^*$ for $\hat \delta^{-1}$=1.0.
As shown in Fig. \ref{fig_sposi_deltas_chis}, the density of bacteria for $\hat \delta^{-1}$=1.0 steeply declines behind the peak of the population wave, and the front side of the wave broadens.
The concentration profile of the nutrient $\hat N$ and the spatial variation of the mean tumbling frequency $<\hat \Psi >$ for $\hat \delta^{-1}$=1.0 are quite different from those in Fig. \ref{fig_psi_NS_delta008}.
In comparison with that in Fig. \ref{fig_psi_NS_delta008}, the gradient of the nutrient $\hat N$ for $\hat \delta^{-1}$=1.0 is rather flat behind the population wave, i.e., $\hat x^* \lesssim -0.2$, and is much steeper in the vicinity of the peak of the population wave, $\hat x^* \simeq 0$.
The gradient of the chemoattractant $\hat S$ is also larger behind the peak of the population wave, $\hat x^*<0$.
These profiles of chemical cues generate a local peak in the modulation of the mean tumbling frequency $<\hat \Psi>$ behind the peak of the population wave.

\begin{figure}[h]
	\centering
	\includegraphics[scale=1.0]{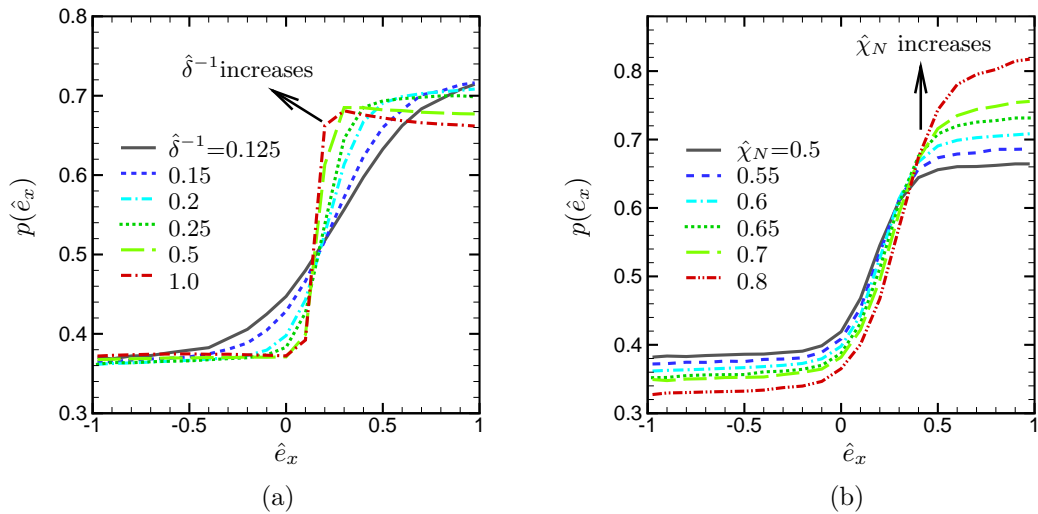}
	\caption{The probability density functions of the longitudinal velocity of bacterium at the peak of the population wave $\hat x^*$=0 with variations in the stiffness parameter $\hat \delta^{-1}$ (a) and modulation parameter $\hat \chi_N$ (b).}\label{fig_pdfx_deltas_chis}
\end{figure}

\begin{figure}[h]
	\centering
	\includegraphics[scale=0.9]{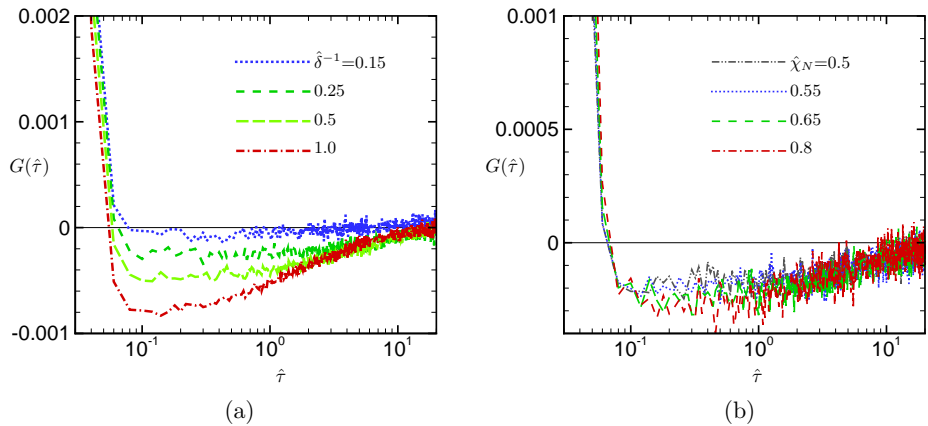}
	\caption{
		The autocorrelation function $G(\hat \tau)$ of the longitudinal relative velocity of the bacteria that form the population wave for different values of the stiffness parameter $\delta^{-1}$ (a) and the modulation parameter (b). See also the caption in Fig. \ref{fig_corr_wv_delta008}.
	}\label{fig_corr_deltas_chis}
\end{figure}

The effects of changing the stiffness and modulation parameters of the response function on the microscopic dynamics of bacterium are shown in Figs. \ref{fig_pdfx_deltas_chis} and \ref{fig_corr_deltas_chis}. 
Figure \ref{fig_pdfx_deltas_chis}(a) shows the effect of changing the stiffness parameter $\hat \delta^{-1}$ of the response function on the PDF of the longitudinal velocity of bacterium. 
As the stiffness parameter increases, the gradients of the PDFs in the vicinity of the traveling speed, $\hat e_x\sim 0.15$, become larger. 
The PDF for the large stiffness parameter, $\hat \delta^{-1}$=0.5 and 1.0, shows an overshooting behavior after the steep gradient in the vicinity of the traveling speed.
This overshoot profile is related to the asymmetric profile of the population density around the peak $\hat x^*\sim 0$ (See Fig. \ref{fig_sposi_deltas_chis}) and the stepwise response function due to the large stiffness parameter and steep gradient of the nutrient around the peak.
An intuitive explanation of the overshoot profile is also given in \ref{appendix_over}.
For a rigorous mathematical description, one can refer Ref. \cite{art:16C}.

In contrast, changing the modulation parameter $\hat \chi_N$ does not affect the gradient of the PDF in the region around the traveling speed, but, at the left and right sides of the steep gradient region, the value of the PDF decreases and increases, respectively, as the modulation parameter increases. (See Fig. \ref{fig_pdfx_deltas_chis}(b).)
The variation of the PDF at large longitudinal velocities due to changes in the modulation parameter is larger than that at small longitudinal velocities, so the mean velocity increases as the modulation parameter increases.

Figure \ref{fig_corr_deltas_chis} shows the ACF of the deviation velocity in the longitudinal direction of the bacteria that form the traveling population wave for various values of the stiffness parameter of the response function.
  The ACF is only slightly affected by the modulation parameter but is significantly affected by the stiffness parameter.
  This fact indicates that the ACF, which represents the microscopic motions of bacteria in the traveling population wave, is highly related to the profile of the population wave. (See also Fig. \ref{fig_sposi_deltas_chis}.)
  As the stiffness parameter increases, the amplitude of the negative correlation, which arises just after an exponential decline due to the Poisson tumbling process, becomes larger. 
  It seems that this behavior is related to the decline of the traveling speed at the large stiffness parameters shown in Fig. \ref{fig_tvel}. 

In summary of this subsection, the stiffness parameter $\hat \delta^{-1}$ of the response function, which represents the sensitivity of a bacterium to chemical cues, significantly affects the wave profile as well as the microscopic motions of bacteria. The traveling speed is only slightly affected by the stiffness parameter.
In contrast, the modulation parameter $\hat \chi_N$ is linearly related to the traveling speed but does not have a significant relationship with the wave profile or the symmetry of the velocity distribution function of the bacterium.

\subsection{Accuracy test}
The accuracy tests are conducted for various numerical parameter settings listed in Table \ref{t_CPS}.
The STD corresponds to that listed in Table \ref{t_parameters} and is used in the results shown in the former subsections.
The T1--T3, M1--M3, and W1--W3 are used to investigate the effects of the variations in time-step size $\Delta \hat t$, mesh interval $\Delta \hat x$, and weight of a simulation particle $w_0$, respectively, on the convergence of density profile $\hat \rho(\hat x)$ and traveling speed $\hat V_{\rm wave}$.
The errors of the density profile and traveling speed are estimated as, respectively,
\begin{equation}\label{err_rho}
	{\rm Err}_{\hat \rho}=\frac{1}{\hat L}\int_0^{\hat L}|\hat \rho'(\hat x)-\hat \rho''(\hat x)|{\rm d}\hat x,
\end{equation}
and
\begin{equation}\label{err_v}
	{\rm Err}_{\hat V_{\rm wave}}=\frac{|\hat V'_{\rm wave} - \hat V''_{\rm wave}|}{V''_{\rm wave}},
\end{equation}
where the prime ``{$\,'\,$}'' and double prime ``{\,$''$\,}''marks indicate the coarser and finer numerical parameters, respectively.

\begin{table}[htbp] 
\centering
\caption{The numerical parameter settings used for the accuracy tests.}\label{t_CPS}
\begin{tabular}{c ccc}
	\hline\hline
	& $\Delta t$ & $\Delta x$ &$w_0$ \\
	\hline
	STD & 0.005 &0.025 & 0.1 \\
	T1 & 0.075 & -- & -- \\
	T2 & 0.0025 & -- & -- \\
	T3 & 0.001 & -- & -- \\
	M1 & -- & 0.1 & -- \\
	M2 & -- & 0.05 & -- \\
	M3 & -- & 0.0125 & -- \\
	W1 & -- & -- & 4.0 \\
	W2 & -- & -- & 1.0 \\
	W3 & -- & -- & 0.05 \\
	\hline\hline
\end{tabular}
\\{The bar ``--'' indicates the same value as for the STD. 
}
\end{table}

\begin{figure}[htbp]
	\centering
	\includegraphics[scale=0.85]{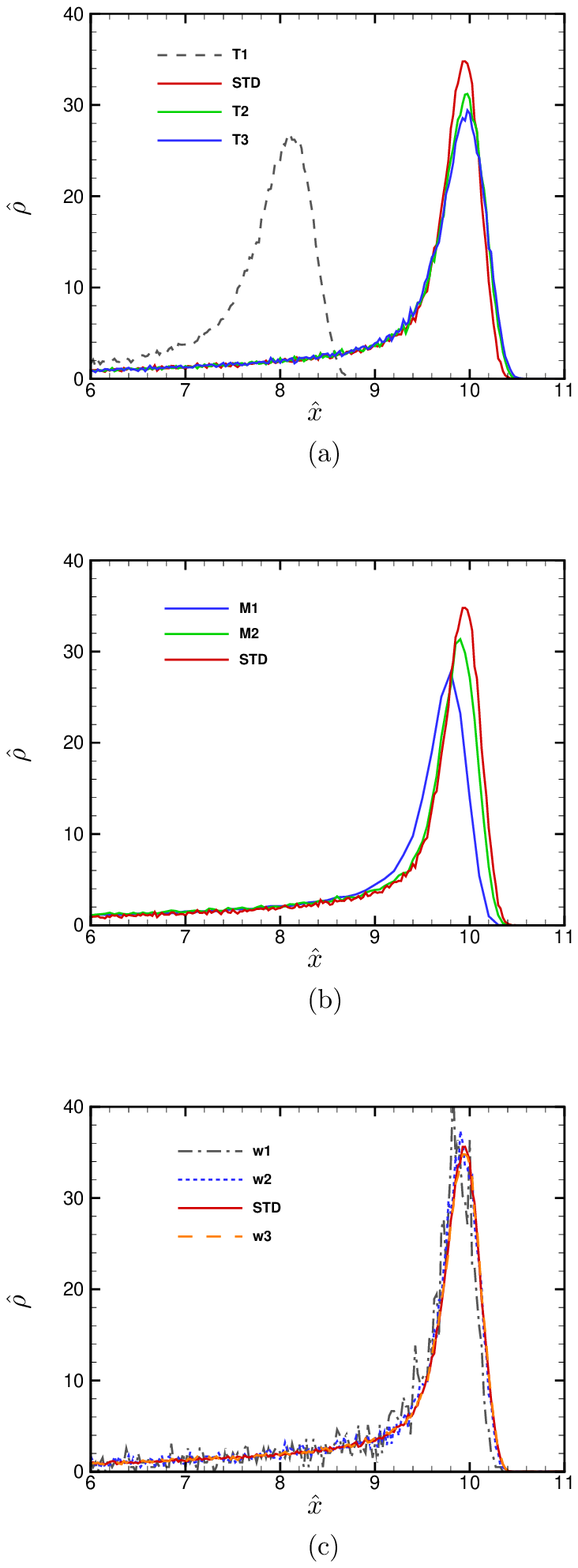}
	\caption{The snapshots of the density profile at $\hat t=60$ obtained for the different numerical parameters in Table \ref{t_CPS}.}\label{fig_dens_diff} 
\end{figure}

Figure \ref{fig_dens_diff} shows the snapshots of the density profiles at $\hat t=60$ obtained for different parameter settings listed in Table \ref{t_CPS}. 
Table \ref{t_EES} shows the results of the error estimates Eqs. (\ref{err_rho}) and (\ref{err_v}). 
\begin{table}[htbp]
	\centering
	\caption{Error estimates between different numerical parameter sets.}\label{t_EES}
	\begin{tabular}{ccc }
		\hline\hline
		$\Delta \hat t$ &${\rm Err}_{\hat \rho}$&${\rm Err}_{\hat V_{\rm wave}}$\\
	      \hline
	      STD--T3 & 0.23 & $3.8\times 10^{-4}$  \\
	      T2---T3 & 0.12 & $2.3\times 10^{-4}$  \\
	 	\multicolumn{3}{c}{(T1 gives a completely different profiles.)}     \\
		\\
		$\Delta \hat x$ &${\rm Err}_{\hat \rho}$&${\rm Err}_{\hat V_{\rm wave}}$\\
	      \hline
	      M1--STD & 0.65 & $3.0\times 10^{-2}$\\
	      M2--STD & 0.24 & $8.1\times 10^{-3}$\\
	   \multicolumn{3}{c}{(M3 induces a numerical divergence.)}    \\
	   \\
               $w_0$           &${\rm Err}_{\hat \rho}$&${\rm Err}_{\hat V_{\rm wave}}$\\
	       \hline
	       W1--W3 & 0.51 & $3.8\times 10^{-3}$ \\
	       W2--W3 & 0.21 & $1.9\times 10^{-3}$ \\
	       STD--W3& 0.074 & $3.4\times 10^{-4}$ \\
		\hline\hline
	\end{tabular}
\end{table}

Except for T1, the variation in time-step size $\Delta \hat t$ affects the density profile only in the vicinity of the peak but less affects the traveling speed.
However, only T1 gives a completely different solution.
This is because for T1, the product of the time-step size and tumbling frequency $\hat \psi_0\Delta \hat t\hat \Psi(\hat {\bm e}_{(l)})$ may exceed the unity.
In the present Monte Carlo method, the tumbling of each simulation particle, say the $l$th particle, is stochastically determined with a probability $\hat \psi_0\Delta \hat t\hat \Psi(\hat {\bm e}_{(l)})$. 
Thus, the time-step size $\Delta \hat t$ must be smaller than the inverse of the maxium tumbling frequency $\hat \psi_{\rm Max}$, which can be estimated as
\begin{equation}\label{eq_maxpsi}
	\hat \psi_{\rm Max}=\hat \psi_0\times\left(1+\frac{\hat \chi_S+\hat \chi_N}{2}\right).
\end{equation}
The mesh interval $\Delta \hat x$ affects both of the density profile and traveling speed.
The computation with M3 setting induces a numerical divergence because this numerical parameter setting does not satisfy the diffusion condition of the transport equations of chemical cues, $\Delta \hat t < \frac{\Delta \hat x^2}{2\hat D_{S,N}}$.
Thus, it is important to note that there are two critical conditions for the time-step size $\Delta \hat t$ and mesh interval $\Delta \hat x$ to be satisfied, i.e.,
\begin{equation}
	\Delta \hat t < \frac{1}{\hat \psi_{\rm Max}},
\end{equation}
\begin{equation}
	\Delta \hat t< \frac{\Delta \hat x^2}{2 \hat D_{S,N}}.
\end{equation}
If the above conditions are satisfied, the numerical solution converges as setting the numerical parameters finer.

The weight parameter $w_0$ (or the number of simulation particles) concerns the fluctuation in the instantaneous density profile, but the traveling speed is less affected by the weight parameter. 
The amplitude of fluctuation measured by Eq. (\ref{err_rho}) decreases approximately in proportion to the square root of weight parameter $w_0^{1/2}$ although the computational cost becomes larger inversely proportional to the weight parameter $w_0$.

\section{Comparison with asymptotic solution}
In this section, the results of the Monte Carlo simulations are compared with the macroscopic transport equation of the population density of bacteria, which is obtained by the asymptotic analysis of the kinetic transport equation.
Here, we only consider the case of the uniform scattering kernel, $\hat K=1/(4\pi)$, and nonproliferation, $\hat r=0$, for the kinetic transport equation.

By introducing a new characteristic time $t_0'$ defined as $t_0'=\psi_0L_0^2/V_0^2$, the kinetic transport equation Eq. (\ref{neq_f}) for the uniform scattering kernel and nonproliferation is rescaled as
\begin{equation}\label{neq_kinetic2}
	\epsilon \frac{\partial \hat f}{\partial \tilde t}+\hat e_\alpha
	\frac{\partial \hat f}{\partial \hat x_\alpha}
	=\frac{1}{4\pi\epsilon}\left\{
		\int_{{\rm all}\,\, \hat {\bm e}'} \hat \Psi(\hat {\bm e}')\hat f(\tilde t, \hat {\bm x}, \hat {\bm e}')d\Omega(\hat {\bm e}')-4\pi \hat \Psi(\hat {\bm e})\hat f(\tilde t,\hat {\bm x},\hat {\bm e})
	\right\},
\end{equation}
where $\epsilon$ corresponds to the Knudsen number, which is defined as the ratio of the mean free path of bacterium $V_0/\psi_0$ to the characteristic length $L_0$ of the system and written as the inverse of the non-dimensional parameter of the mean tumbling frequency, i.e., $\epsilon=\hat \psi_0^{-1}$. (See also Sec. \ref{subsec_non_dim}.) 
Hereafter, the superscript ``tilde'', e.g., $\tilde t$, represents the non-dimensional variable rescaled with the characteristic time $t_0'$.

Suppose that the modulations $\hat \chi_{N}$ and $\hat \chi_{S}$ are of amplitude $\epsilon$, respectively, i.e., $\hat \chi_{N,S}=\epsilon \phi_{N,S}$ in Eq. (\ref{neq_psi2}), 
and consider the asymptotic limit (or the continuum limit) $\epsilon \rightarrow 0$ for Eq. (\ref{neq_kinetic2}).
Then, we obtain the following drift-diffusion equation for the population density of bacteria $\hat \rho(\tilde t,\hat {\bm x})$,\cite{art:10SCBBSP} (See also Appendix)
\begin{equation}\label{eq_asymprho}
	\frac{\partial \hat \rho}{\partial \tilde t}=\tilde D_{\rho}
	\frac{\partial^2 \hat \rho}{\partial \hat x_\alpha^2}+
	\frac{\partial}{\partial \hat x_\alpha}\left[
		\hat \rho (u_\alpha[\log \tilde S]+u_\alpha[\log \hat N])
	\right],
\end{equation}
where the flux $u_\alpha$ is the functional of the gradients of chemical cues and defined as
\begin{equation}\label{eq_flux}
	u_\alpha[Y]=-\frac{\phi_Y}{4}\frac{\nabla Y}{|\nabla Y|}
	\int_{-1}^1\xi \tanh\left(\frac{1}{\hat \delta}|\nabla Y|\xi\right)d\xi.
\end{equation}
The velocity distribution of bacteria becomes uniform in the continuum limit $\epsilon \rightarrow 0$, i.e., $\hat f(t,\hat {\bm x},\hat {\bm e})= \hat \rho(\hat t,\hat {\bm x})/4\pi$.
Note that the reaction-diffusion equations of chemical cues Eqs. (\ref{neq_S}) and (\ref{neq_N}) are also rescaled by the characteristic time $t_0'$.
The relations between the non-dimensional forms for the diffusion coefficients of chemical cues $D_{S, N}$, the production rate of attractant $a$, the consumption rate of nutrient $c$, and time $t$ scaled by $t_0'$ (described by the superscript ``tilde'') and $t_0$ (described by the superscript ``hat''), respectively, are written as 
\begin{gather}\label{relation_nonpara}
	\tilde D_{N,S}=\hat D_{N,S}/\epsilon,\quad \tilde a=\hat a/\epsilon,\quad \tilde c=\hat c/\epsilon, \quad \tilde t=\epsilon \hat t.
\end{gather}

The Monte Carlo simulations are conducted with variations in the Knudsen number $\epsilon$, i.e., $\epsilon$=0.02, 0.01, 0.005, 0.002, and 0.001, whereas the values of the other non-dimensional parameters are fixed as shown in Table \ref{t_parameters_asympt}.
\begin{table}[htbp]
	\centering
	\caption{The values of non-dimensional parameters for asymptotic equations.}\label{t_parameters_asympt}
	\begin{tabular}{lc}
		\hline\hline
		degradation rate of chemoattractant $\tilde a$ & 24.0\\
		production rate of chemoattractant $\tilde b$ &1.0 \\
		consumption rate of nutrient $\tilde c$& 120.0\\
		diffusion coefficient of nutrient $\tilde D_N$& 3.84 \\
		diffusion coefficient of chemoattractant $\tilde D_S$ &3.84 \\
		modulation of tumble freq. $\phi_N$  & 72.0 \\
		modulation of tumble freq. $\phi_S$  &  24.0 \\
		stiffness of the response function $\hat \delta^{-1}$ &0.2\\
		\hline\hline
	\end{tabular}
\end{table}
These values are set to coincide with those listed in Table \ref{t_parameters} at $\hat \psi_0=120$.
The drift-diffusion equation for the population density of bacteria $\hat \rho$, Eq. (\ref{eq_asymprho}), coupled with Eqs. (\ref{neq_S}) and (\ref{neq_N}), is numerically calculated with the finite volume method for the parameters listed in Table \ref{t_parameters_asympt}.
The flux $u_\alpha$ is calculated at each time step with using the Simpson's integral method according to the gradients of chemical cues obtained at the previous time step.
To obtain accurate results for small Knudsen numbers in the Monte Carlo simulations, which are compared with the asymptotic solution in the continuum limit, the time-step size and particle number are set as $\Delta \hat t=1 \times 10^{-4}$ and $M_0=226560$ (or $w_0=0.025$), respectively, in the Monte Carlo simulations. 

\begin{figure}[h]
	\centering
	\includegraphics[scale=1.0]{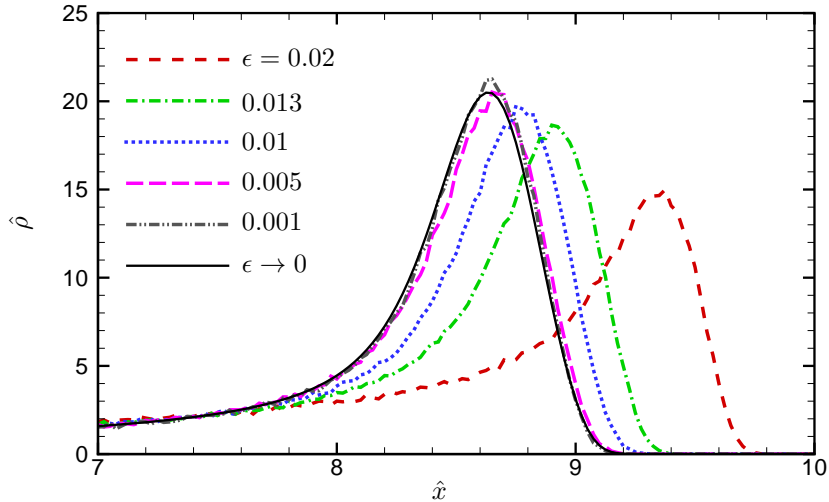}
	\caption{Comparison of the population densities of bacteria for different Knudsen numbers and for the asymptotic limit at a time $\tilde t=0.5$.}\label{fig_comp_asympt}
\end{figure}

Figure \ref{fig_comp_asympt} shows the snapshots of the population density of bacterial $\hat \rho(\hat x)$ at $\tilde t=0.5$ for various Knudsen numbers, i.e., $\epsilon$=0.02, 0.013, 0.01, 0.005, and 0.001, obtained by the Monte Carlo simulations and the snapshot for the  asymptotic limit obtained by the finite volume calculation of Eqs. (\ref{eq_asymprho}) and (\ref{eq_flux}), coupled with Eqs. (\ref{neq_S}) and (\ref{neq_N}).
It is seen that the results of the Monte Carlo simulations are asymptotically close to the asymptotic solution in the continuum limit as the Knudsen number decreases; the snapshot of the population density for $\epsilon=0.001$ obtained by the Monte Carlo simulation almost coincides with the asymptotic solution in the continuum limit.
However, significant deviations from the asymptotic solution are also observed for moderately small Knudsen numbers, i.e., $\epsilon\gtrsim$0.01.
These deviations demonstrate that the asymptotic solution in the continuum limit is no more valid for that moderately-small Knudsen number.
\begin{figure}[h]
	\centering
	\includegraphics[scale=1.0]{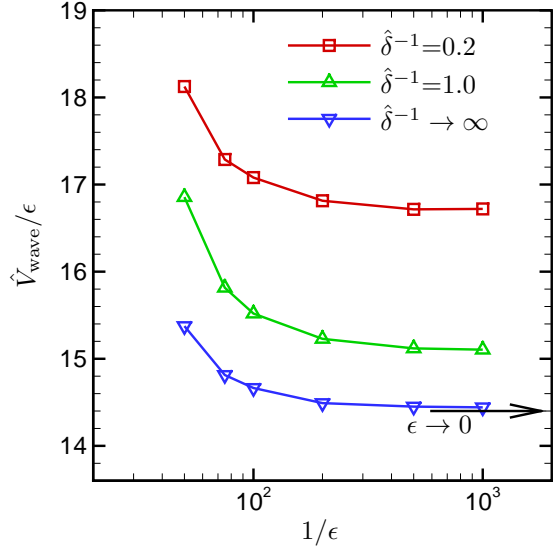}
	\caption{The traveling speed of the population wave vs. the inverse of the Knudsen number for different stiffness parameters. For $\delta^{-1}\rightarrow\infty$, the hyperbolic tangent in Eq. (\ref{neq_psi2}) is replaced with the sign function $\hat X/|\hat X|$. The left arrow shows the traveling speed for $\delta^{-1}\rightarrow\infty$ in the continuum limit obtained by the analytical formula in Ref. \cite{art:10SCBBSP}.
}\label{fig_tvel_eps}
\end{figure}
Figure \ref{fig_tvel_eps} shows the convergence of the traveling speed $\hat V_{\rm wave}$ scaled by $\epsilon$ in the continuum limit $\epsilon \rightarrow 0$.
It is seen that as increasing the stiffness parameter $\delta^{-1}$, the results obtained with Eq. (\ref{neq_psi2}) converges to those obtained with the sign response function and the result of the sign response function converges to that obtained by the analytical formula in Ref. \cite{art:10SCBBSP} in the continuum limit $\epsilon\rightarrow 0$.
This also demonstrates that the present Monte Carlo method can accurately reproduce the analytical result of the traveling speed obtained in Ref. \cite{art:10SCBBSP}.

\begin{figure}[h]
	\centering
	\includegraphics[scale=0.9]{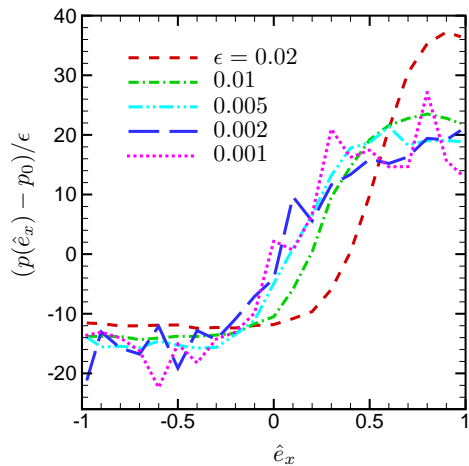}
	\caption{Comparison of the first-order probability density functions of longitudinal velocity for different Knudsen numbers at the peak of the population wave.
	$p_0$ is the uniform distribution, i.e., $p_0=0.5$, and represents the probability density in the continuum limit $\epsilon\rightarrow 0$.
}\label{fig_pdf_asympt}
\end{figure}

Figure \ref{fig_pdf_asympt} shows the deviations of the probability density functions of longitudinal velocity $p(\hat e_x)$ from the uniform probability density $p_0$ scaled by the Knudsen number $\epsilon$ at the peak of population density $\hat x^*=0$, which represents the first order probability density function in the asymptotic expansion of $\epsilon$. 
It seems that the first order probability density asymptotically converges as the Knudsen number $\epsilon$ decreases from 0.02 to 0.002, although the convergence of the first order probability density in the asymptotic limit is not very clearly seen because the fluctuation is notable at $\epsilon=0.001$.
This result also indicates that the probability density $p(\hat e_x)$ converges to the uniform probability density $p_0$ in the continuum limit $\epsilon\rightarrow 0$.

It should be noted that the Knudsen number depends not only on the biological properties of bacterium but also on the characteristic length $L_0$ of the system.
Thus, the Knudsen number becomes significant when one considers, for example, the cluster formation of bacteria in the micro devices and the collective migration of bacteria in the vicinity of a boundary.
The deviations observed for moderately-small Knudsen numbers indicate that the present Monte Carlo method can take on a significance in the investigation on the micro scale systems.

In the summary of this section, the asymptotic behaviors of the population density of bacteria and the velocity distribution of the bacterium are numerically demonstrated and the results obtained by the Monte Carlo simulations for small Knudsen numbers are compared to the asymptotic solution in the continuum limit.
The comparison to the asymptotic solution validates the accuracy of the present Monte Carlo method in the asymptotic behaviors for small Knudsen numbers.
The distinctive deviations of the results obtained by Monte Carlo simulations from the asymptotic solution are observed in both the population density and the velocity distribution of bacteria for moderately small Knudsen numbers $\epsilon\gtrsim$0.01.
This result demonstrates the restriction of the continuum model and the utility of the present Monte Carlo simulation in the investigation on the micro-scale systems with moderately-small Knudsen numbers.

\section{Summary}

In this paper, a Monte Carlo simulation method for chemotactic bacteria is newly developed on the basis of the kinetic chemotaxis model, which was proposed in Ref. \cite{art:11SCBPBS}. 
In this method, the Monte Carlo algorithm is employed to calculate the microscopic motions of bacteria that can be described using a model response function and a scattering kernel.
The response function and scattering kernel depend on the local concentration of chemical cues, which is calculated using a finite-volume method, so the microscopic motions of bacteria are coupled to the macroscopic transport of chemical cues via the response function and scattering kernel.

The present simulation method can successfully reproduce the traveling population wave of chemotactic bacteria in a microchannel (Fig. \ref{fig_twave_sposi}).
The traveling speed and the wave profile obtained by the simulation are close to those obtained in experiments.
The microscopic dynamics of the bacteria that form the traveling population wave are investigated in terms of the probability density function (PDF) and the autocorrelation function (ACF) of the velocity of bacteria (Figs. \ref{fig_pdfxy_delta008} and \ref{fig_corr_wv_delta008}).
The PDF of the lateral velocity of a bacterium is almost constant and uniform, except when the amplitude of the velocity is large.
However, the PDF of the longitudinal velocity monotonically increases as the longitudinal velocity increases, so the PDF is biased toward the positive velocity regime.
The PDF of the longitudinal velocity also spatially changes; the gradient of the PDF in the vicinity of the traveling speed of the population wave increases as the spatial position approaches the peak of the population wave, and the step-function-like profile is obtained at the peak of the population wave.
This stepwise profile is related to the steep spatial gradient of nutrients around the peak of the population wave.

The ACF of the deviation velocity, which is calculated relative to the mean velocity, in the longitudinal direction shows a rapid exponential decline on the short time scale due to the Poisson tumbling process as well as the negative correlation that occurs after the Poisson tumbling process, which lasts a long time period.
The negative autocorrelation function characterizes the random motions of the bacteria trapped within the population wave. 
The time period of the negative autocorrelation can be estimated as the characteristic time in which the traveling population wave passes through the width of density profile. 

The response function, which determines the microscopic behavior of bacteria, combines two important parameters, that is, the stiffness parameter, which represents the sensitivity of bacteria, and the modulation parameter, which describes the biased movements of bacteria toward chemical cues.
The effect of changing the parameters of the response function on the macroscopic transport and the microscopic dynamics of bacteria is also investigated.
The results showed that the stiffness parameter significantly affects both the wave profile and the microscopic motions of bacteria. The traveling speed is only slightly affected by the stiffness parameter.
In contrast, the modulation parameter $\hat \chi_N$ is linearly related to the traveling speed but does not significantly affect the wave profile or the microscopic motions of bacteria.

The numerical accuracy of the method is tested with variation in the time-step size $\Delta t$, mesh-interval $\Delta x$, and weight parameter $w_0$.
It is found that the time-step size must be smaller than the minimum mean run duration, i.e., $\Delta \hat t<\hat \psi_{\rm Max}^{-1}$. Otherwise the Monte Carlo simulation produces a physically incompatible result. The time-step size must satisfy the diffusion conditions of chemical cues, i.e., $\Delta \hat t<\Delta \hat x^2/(2\hat D_{S,N})$. 
The weight parameter affects the fluctuation in the instantaneous density profile but less affects the traveling speed. 
The fluctuation decreases as decreasing the weight parameter (or increasing the particle number).

The drift-diffusion equation for the bacteria population density is derived by the asymptotic analysis of the kinetic chemotaxis equation in the continuum limit, where the Knudsen number $\epsilon$, which is defined by the ratio of the mean free path of bacterium to the characteristic length of the system, vanishes.
In the present study, the results of Monte Carlo simulations for small Knudsen numbers are numerically compared with the asymptotic solution.
It is demonstrated that the results of the Monte Carlo simulations for very small Knudsen numbers, i.e., $\epsilon\lesssim 0.005$, almost coincide with the asymptotic solution in the continuum limit but the deviation becomes notable for $\epsilon \gtrsim 0.01$.
Thus, it is concluded that the continuum equation is valid for only a very small Knudsen number, e.g., $\epsilon\lesssim 0.005$, but for a finite value of the Kundsen number, say $\epsilon\gtrsim 0.01$, one needs to utilize the kinetic chemotaxis model.
The comparison of the Monte Carlo simulation and the asymptotic solution also demonstrates the validity of the present method in the asymptotic behaviors of both the macroscopic transport of bacteria population density and the microscopic dynamics of the bacterium.

It should be noted that the present kinetic chemotaxis model cannot treat the multi-body steric interactions and long-range hydrodynamic interactions between the bacteria, although those are supposed to be significant in a highly concentrated population. 
However, the kinetic chemotaxis model takes on great significance in the investigation of the multiscale mechanism inherent in the chemotaxis problems in a moderate concentration where the steric and hydrodynamic effects can be ignored.

In the present study, the Monte Carlo simulation is applied to a simple and fundamental problem of chemotactic bacteria in order to examine the validity of the present method.
The method also succeeds in numerically demonstrating an asymptotic relationship between a continuum model and a kinetic model for the present problem.
There remain two important challenges to extending the present Monte Carlo method for applications to more complicated transport phenomena involved in practical engineering and biological systems.
First, the extension is of the geometry of the problem.
The present Monte Carlo method treats the velocity space in the three-dimensional Cartesian coordinate, but the spatial geometry is restricted in the one-dimensional system.
It is important to extend the method to the multi-dimensional space problem and demonstrate the colony dynamics in multi-dimensional spaces.
Second, the extension is of the response function to involve the memory effect of bacterium along its pathway.
This extension allows us to investigate the memory effects of bacterium in the colony dynamics.\cite{art:04G,art:05CG,art:12NNS,art:15PTV}
The particle-based algorithm employed in the present Monte Carlo method is thought to be useful in incorporating the memories of each bacterium into the present Monte Carlo method.
These extensions are important future works.

\section*{Acknowledgements} 
The author expresses his sincere gratitude to Vincent CALVEZ, Francis FILBET, Beno\^it PERTHAME, and Kazuo AOKI for useful discussions and helpful suggestions.
This study was financially supported by JSPS KAKENHI Grant Number 26247069 and 15KT0110 and the Grant from CASIO SCIENCE PROMOTION FOUNDATION.

\appendix

\section{Asymptotic analysis}
In this appendix, the macroscopic transport equation of the population density of bacteria is derived by the asymptotic analysis of the kinetic chemotaxis model Eq. (\ref{neq_kinetic2}) according to Ref. \cite{art:10SCBBSP}.
The drift diffusion equation of the population density of bacteria is obtained in the continuum limit, where the Knudsen number $\epsilon$, which is defined by the ratio of the mean free path of bacterium $V_0/\psi_0$ to the characteristic length $L_0$ of the system, i.e., $\epsilon=V_0/\psi_0L_0$, vanishes.
The basic procedure of the asymptotic analysis was put forward in Ref. \cite{art:10SCBBSP}, but the functional form of the response function of bacteria used in the present study is only different from that in the reference.

We seek the asymptotic solution of the bacterial density $\hat f(\tilde t,\hat {\bm x},\hat {\bm e})$ in a expansion of $\epsilon$, i.e., $\hat f=\hat f_0+\epsilon \hat f_1+\epsilon \hat f_2+\cdots$. By substituting the expansion form into Eq. (\ref{neq_kinetic2}) and arranging the terms by the order in $\epsilon$, we obtain the following expansion,
\begin{align}\label{exp_kinetic}
	&\hat e_\alpha \frac{\partial \hat f_0}{\partial \hat x_\alpha}
	+\epsilon\left(\frac{\partial \hat f_0}{\partial \tilde t}
	+\hat e_\alpha\frac{\partial \hat f_1}{\partial \hat x_\alpha}\right)
	\nonumber \\
	&=\frac{1}{4\pi\epsilon}\left\{
		\left[ \int_{{\rm all}\, \hat {\bm e}'}\hat f_0(\hat {\bm e}') d\Omega(\hat {\bm e}')
		-4\pi\hat f_0(\hat {\bm e})\right ]
		\right.
		\nonumber \\
		&+\epsilon\left [
			\int_{{\rm all}\,\hat {\bm e}'}\
			(\hat \Psi_1(\hat {\bm e}')\hat f_0(\hat {\bm e}')
			+\hat f_1(\hat {\bm e}'))
			d\Omega(\hat {\bm e}')
			-4\pi\hat f_1(\hat {\bm e})
			+4\pi \hat \Psi_1(\hat {\bm e})\hat f_0(\hat {\bm e})
		\right]
		\nonumber \\
		&+{\cal O}(\epsilon^2)
	\left.\right\},
\end{align}
where $\hat \Psi_1(\hat {\bm e})$ is obtained by the expansion of Eq. (\ref{neq_Psi}) and is written as
\begin{equation}\label{exp_Psi}
	\hat \Psi_1(\hat {\bm e})=\frac{\phi_S}{2}\tanh\left(\frac{\hat {\bm e}\cdot\nabla \log S}{\hat \delta}\right)
	+\frac{\phi_N}{2}\tanh\left(\frac{\hat {\bm e}\cdot\nabla \log N}{\hat \delta}\right).
\end{equation}
Here, we suppose that the modulation parameters $\hat \chi_S$ and $\hat \chi_N$ are of amplitude $\epsilon$, respectively, i.e., $\hat \chi_{S,N}=\epsilon\phi_{S,N}$.

For the limit $\epsilon \rightarrow 0$, we obtain an uniform distribution as to the bacterial velocity, i.e.,
\begin{equation}\label{eq_f0}
	\hat f_0(\tilde t,\hat {\bm x}, \hat {\bm e})
	=\frac{\hat \rho_0(\tilde t,\hat {\bm x})}{4\pi},
\end{equation}
where $\hat \rho_0(\tilde t,\hat {\bm x})$ is the population density of bacteria in the asymptotic limit and is obtained by $\hat \rho_0=\int \hat f_0(\hat {\bm e})d\Omega(\hat {\bm e})$.
Hereafter, unless otherwise stated, the integral as to the vector $\hat {\bm e}$ is performed over all directions of $\hat {\bm e}$.
By substituting Eq. (\ref{eq_f0}) into Eq. (\ref{exp_kinetic}) and integrating Eq. (\ref{exp_kinetic}) as to $\hat {\bm e}$, we obtain the following equation for $\hat \rho_0$,
\begin{equation}\label{eq_rho0}
	\frac{\partial \hat \rho_0}{\partial \tilde t}+\nabla\cdot \hat {\bm j}_1=0,
\end{equation}
where the flux $\hat {\bm j}_1$ is defined as $\hat {\bm j}_1=\int\hat {\bm e}\hat f_1(\hat {\bm e})d\Omega(\hat {\bm e})$.
By integrating Eq. (\ref{exp_kinetic}) multiplied by $\hat {\bm e}$ as to $\hat {\bm e}$, we obtain the equation for the flux $\hat {\bm j}_1$ in the first order of $\epsilon$,
\begin{equation} \label{eq_j1}
	\hat j_{1\,\alpha}=
	-\frac{1}{4\pi}\left(
	\int \hat e_\alpha \hat e_\beta d\Omega(\hat {\bm e}) 
	\right)
	\frac{\partial \hat \rho_0}{\partial \hat x_\beta}
	+\frac{1}{4\pi}
	\hat\rho_0\int \hat e_\alpha\Psi_1(\hat {\bm e})d\Omega(\hat {\bm e}).
\end{equation}
The integral in the first term of the right hand side of Eq. (\ref{eq_j1}) is calculated as $\int \hat e_\alpha \hat e_\beta d\Omega(\hat {\bm e})=(4\pi/3)\delta_{\alpha\beta}$.
The integrals of the first and second terms of Eq. (\ref{exp_Psi}) , which are involved in the second term of the right hand side of Eq. (\ref{eq_j1}), are calculated by using the polar coordinates whose pole directions are set to the directions of gradient vector $\nabla \log S$ and $\nabla \log N$, respectively.
Thus, we obtain the equation for the flux $\hat {\bm j}_1$,
\begin{align}\label{eq_j1_2}
	\hat {\bm j}_1=-\tilde D_\rho \nabla \hat \rho_0
	+\hat \rho_0
	&\left[
		\frac{\phi_S}{4}\frac{\nabla \log S}{|\nabla \log S|}
		\int_{-1}^1 \xi \tanh\left(
		\frac{1}{\hat \delta}|\nabla \log S|\xi
		\right)
		d\xi 
		\right.
		\nonumber \\
		&\left .
		+\frac{\phi_N}{4}\frac{\nabla \log N}{|\nabla \log N|}
		\int_{-1}^1 \xi \tanh\left(
		\frac{1}{\hat \delta}|\nabla \log N|\xi
		\right)
		d\xi
	\right],
\end{align}
where $\tilde D_\rho=1/3$.
By substituting Eq. (\ref{eq_j1_2}) into Eq. (\ref{eq_rho0}), we obtain the drift diffusion equation for the population density in the asymptotic limit.

\section{Overshoot profile of the velocity distribution}
\label{appendix_over}
In this appendix, we discuss the overshoot profile of the velocity distribution which appears at the peak of the population wave for a large stiffness parameter $\hat \delta^{-1}$. See Fig. \ref{fig_pdfx_deltas_chis}.
For the simplicity, we consider the uniform scattering kernel $\hat K=1/4\pi$ and nonproliferation $\hat r$=0 because these factors less affects the results.
In the one-dimensional relative coordinate system as introduced in Eq. (\ref{eq_xrc}), the density of the bacteria with a velocity $\hat e_x$, $\hat f(\hat x^*,\hat e_x)$ can be written as 
\begin{equation}\label{eq_f_rc} 
	\hat f(\hat x^*,\hat e_x)=\frac{G}{\hat \lambda(\hat e_x)}-
	\left(\frac{\hat e_x-\hat V_{\rm wave}}{\hat \lambda(\hat e_x)}\right)
	\frac{\partial \hat f}{\partial \hat x^*},
\end{equation}
where $G$ is the generation rate due to the tumbling and is written as
\begin{equation}\label{eq_gene} 
	G=\frac{1}{2}\int_{-1}^1\hat \lambda(\hat e_x)\hat f(\hat x^*,\hat e_x)d\hat e_x,
\end{equation}
and the tumbling rate $\hat \lambda(\hat e_x)$ is written as
\begin{equation}
	\hat \lambda(\hat e_x)=\hat \psi_0\left[
		1-\frac{\hat \chi_N}{2}
		\tanh\left(\frac{\hat e_x-\hat V_{\rm wave}}{\hat \delta}		\frac{\partial \log{\hat N}}{\partial \hat x^*}\right)
		-\frac{\hat \chi_S}{2}
		\tanh\left(\frac{\hat e_x-\hat V_{\rm wave}}{\hat \delta}
		\frac{\partial \log{\hat S}}{\partial \hat x^*}\right)
	\right].
\end{equation}
Here, we assume the velocity space for $\hat {\bm e}$ is axially symmetric and $\partial /\partial y=\partial /\partial z=0$.
Note that the generation rate $G$ is independent of $\hat e_x$ and
$\hat \lambda(\hat e_x)^{-1}$ represents the mean run duration of the bacteria with a velocity $\hat e_x$.
The mean run distance of the bacteria with a velocity $\hat e_x$ in the relative coordinate system can be written as
\begin{equation}
	\hat l^*(\hat e_x)=\frac{\hat e_x-\hat V_{\rm wave}}{\hat \lambda(\hat e_x)}.
\end{equation}
The density of the bacteria with a velocity $\hat e_x$ at the peak of the wave $\hat x^*=0$ can be approximately written as, for a small $\hat l^*$,
\begin{equation}\label{eq_f_rc0}
\hat f(0,\hat e_x)=\frac{G}{\hat \lambda(\hat e_x)}+
\frac{1}{2}\left[
	\hat f(-\hat l^*,\hat e_x)-\hat f(\hat l^*,\hat e_x)
\right].
\end{equation}
Here, the first term of the right-hand side of Eq. (\ref{eq_f_rc0}), which we call the generation term, represents the generation of a new velocity $\hat e_x$ due to the tumbling in a mean run duration $\hat \lambda(\hat e_x)^{-1}$ and the second term of the right-hand side of Eq. (\ref{eq_f_rc0}), which we call the flux term, represents the net flux of the density of the bacteria migrating without tumbling.

The tumbling rate at the peak of the wave $\hat \lambda(\hat e_x)$ steeply decreases in the vicinity of $\hat e_x=\hat V_{\rm wave}$ because the spatial gradient of nutrient concentration is very large at the peak of the wave.
Thus, the velocity distribution $\hat f(0,\hat e_x)$ steeply increases in the vicinity of $\hat e_x=\hat V_{\rm wave}$ due to the contribution of the generation term in Eq. (\ref{eq_f_rc0}) while the contribution of the flux term is negligible because the mean run distance vanishes at $\hat e_x=V_{\rm wave}$. 
However, the flux term becomes significant as the velocity $\hat e_x$ separates from the traveling speed $\hat V_{\rm wave}$.
Especially, for a large stiffness parameter $\hat \delta^{-1}$, the tumbling rate $\hat \lambda(\hat e_x)$ is a stepwise function of $\hat e_x - V_{\rm wave}$, so that, except only the vicinity of $\hat e_x=V_{\rm wave}$, the mean run distance $\hat l^*(\hat e_x)$ is proportional to $\hat e_x - V_{\rm wave}$ while the generation term does not vary significantly.
In addition, the density profile around the peak of the wave becomes highly asymmetric as the stiffness parameter $\hat \delta^{-1}$ is large; the rear side of the wave steeply declines while the front side of the wave broadens.  
Thus, for $\hat e_x-V_{\rm wave}>0$, the density at the front side of the peak $\hat f(\hat l^*,\hat e_x)$ is larger than that at the rear side of the peak $\hat f(-\hat l^*,\hat e_x)$, and the flux term of Eq. (\ref{eq_f_rc0}) negatively grows as $\hat e_x-V_{\rm wave}$.
This gives the overshoot profile of the velocity distribution at the peak of the wave for a large stiffness parameter. 

\end{document}